\documentclass[twocolumn, trackchanges]{aastex631}
\usepackage{color,comment}
\hypersetup{linkcolor=blue,citecolor=blue,urlcolor=blue}

\newcommand{\secref}[1]{\S\ref{#1}}
\graphicspath{{./}{figures/}}
\usepackage{float}
\usepackage{amsmath}
\shortauthors{Garg et al.}
\begin{document}

\title{The BPT Diagram in Cosmological Galaxy Formation Simulations: Understanding the Physics Driving Offsets at High-Redshift}

\correspondingauthor{Prerak Garg}
\email{prerakgarg@ufl.edu}

\author[0000-0002-5923-2151]{Prerak Garg}
\affil{Department of Astronomy, University of Florida, 211 Bryant Space Sciences Center, Gainesville, FL 32611, USA}

\author[0000-0002-7064-4309]{Desika Narayanan}
\affil{Department of Astronomy, University of Florida, 211 Bryant Space Sciences Center, Gainesville, FL 32611, USA}
\affil{University of Florida Informatics Institute, 432 Newell Drive, CISE Bldg E251, Gainesville, FL 32611, USA}
\affil{Cosmic Dawn Center at the Niels Bohr Institute, University of Copenhagen and DTU-Space, Technical University of Denmark, Copenhagen, Denmark}

\author[0000-0002-7392-3637]{Nell Byler}
\affil{Pacific Northwest National Laboratory, 902 Battelle Boulevard, Richland, WA 99352, USA}

\author[0000-0003-4792-9119]{Ryan L. Sanders}
\affil{Department of Physics, University of California, Davis, One Shields Avenue, Davis, CA 95616, USA}

\author[0000-0003-3509-4855]{Alice E. Shapley}
\affil{Department of Physics and Astronomy, University of California, Los Angeles, 430 Portola Plaza, Los Angeles, CA 90095, USA}

\author[0000-0001-6369-1636]{Allison L. Strom}
\affil{Department of Astrophysical Sciences, 4 Ivy Lane, Princeton University, Princeton, NJ 08544, USA}

\author[0000-0003-2842-9434]{Romeel Dav\'e}
\affil{Institute for Astronomy, Royal Observatory Edinburgh, EH9 3HJ, UK}
\affil{University of the Western Cape, Bellville, Cape Town 7535, South Africa}
\affil{South African Astronomical Observatory, Cape Town 7925, South Africa}

\author[0000-0002-3301-3321]{Michaela Hirschmann}
\affil{DARK, Niels Bohr Institute, University of Copenhagen, Niels Bohr
Building, Jagtvej 128, DK-2200 Copenhagen, Denmark}

\author[0000-0001-7964-5933]{Christopher C. Lovell}
\affil{Centre for Astrophysical Research, School of Physics, Astronomy and Mathematics, University of Hertfordshire, Hatfield, AL10 9AB, UK}

\author[0000-0003-3191-9039]{Justin Otter}
\affil{Department of Physics $\&$ Astronomy, Johns Hopkins University, Bloomberg Center, 3400 N Charles Street., Baltimore, MD 21218, USA}

\author[0000-0003-1151-4659]{Gerg\"o Popping}
\affil{European Southern Observatory, Karl-Schwarzschild-Strasse 2, D-85748, Garching, Germany}

\author[0000-0003-1151-4659]{George C. Privon}
\affil{National Radio Astronomy Observatory, 520 Edgemont Road, Charlottesville, VA, 22903, USA}

\begin{abstract}
The Baldwin, Philips, \& Terlevich diagram of [O\scalebox{0.9}{ III}]/H$\beta$ vs. [N\scalebox{0.9}{ II}]/H$\alpha$ (hereafter N2-BPT) has long been used as a tool for classifying galaxies based on the dominant source of ionizing radiation. Recent observations have demonstrated that galaxies at $z\sim2$ reside offset from local galaxies in the N2-BPT space. In this paper, we conduct a series of controlled numerical experiments to understand the potential physical processes driving this offset. We model nebular line emission in a large sample of galaxies, taken from the \textsc{simba} cosmological hydrodynamic galaxy formation simulation, using the \textsc{cloudy} photoionization code to compute the nebular line luminosities from H\scalebox{0.9}{ II} regions. We find that the observed shift toward higher [O\scalebox{0.9}{ III}]/H$\beta$ and [N\scalebox{0.9}{ II}]/H$\alpha$ values at high redshift arises from sample selection: when we consider only the most massive galaxies $M_* \sim 10^{10-11} M_\odot$, the offset naturally appears, due to their high metallicities. We predict that deeper observations that probe lower-mass galaxies will reveal galaxies that lie on a locus comparable to $z\sim 0$ observations.  Even when accounting for sample selection effects, we find that there is a subtle mismatch between simulations and observations. To resolve this discrepancy, we investigate the impact of varying ionization parameters, H\scalebox{0.9}{ II} region densities, gas-phase abundance patterns, and increasing radiation field hardness on N2-BPT diagrams. We find that either decreasing the ionization parameter or increasing the N/O ratio of galaxies at fixed O/H can move galaxies along a self-similar arc in N2-BPT space that is occupied by high-redshift galaxies.
\end{abstract}


\section{Introduction} \label{sec:intro}
Understanding the evolution of the physical properties of galaxies over cosmic time is a primary goal of the field of galaxy formation and evolution. Diagnostics based on nebular line emission have played an especially important role in constraining the properties of ionized gas. Nebular lines, first discovered in the late $19^{\rm th}$ century, were shown by \citet{bowen1927} to arise from forbidden transitions of highly ionized metal species. Since nebular emission arises from metals in ionized gas, which is generally found in H\scalebox{0.9}{ II} regions around O and B stars, it is commonly used as a tracer of young stellar populations in galaxies. This makes it a valuable tool to probe galaxy-wide properties like metallicity, star formation rate (SFR), star formation history, etc. \citep{osterbrook2006}.

For example, metallicity calibrations made from small samples of bright nearby objects with auroral line measurements or photoionization models using strong line ratios have become the industry standard.  Temperature-sensitive auroral lines such as [O\scalebox{0.9}{ III}] {$\lambda$4363}, [N\scalebox{0.9}{ II}] {$\lambda$5755}, [S\scalebox{0.9}{ III}] {$\lambda$6312}, and [S\scalebox{0.9}{ II}] {$\lambda$4072} can be used to determine both the electron temperature $(T_e)$\citep[e.g.][]{McLennan1925, Keenan1996} and with an assumption of an ionization correction factor, oxygen abundances, and metallicities \citep{Pagel1992, osterbrook2006, Izotov2006, PrezMontero_2017}. This method provides one of the most accurate measurements of ionized-gas metallicity and has been widely used in local galactic and extragalactic H\scalebox{0.9}{ II} region studies \citep{Bresolin_2007, Monreal2012, Westmoquette2013, andrews2013, Maseda_2014, Berg_2015, McLeod2016, Croxall_2015, Croxall_2016, Pilyugin2016, Lin_2017}. Unfortunately, these auroral lines are quite faint, making it difficult to observe them in faraway galaxies and even in some O-rich local sources. Studies in the past decade have only been able to observe them at $z \gtrapprox 1$ in a handful of galaxies \citep{villar2004, Yuan_2009, erb2010, rigby2011, Brammer_2012, Christensen2012a, Christensen2012b, Stark2014, Bayliss_2014, James2014, Jones2015, steidel2014, Sanders_2016, Kojima2017, Berg_2018, Patricio2018, Sanders_direct2020, Gburek_2019}.

Because of the above-mentioned limitations with using auroral lines, many researchers have turned to using strong line calibrations to derive the physical conditions in ionized gas. Over the years a wide variety of diagnostic calibrations have been constructed to translate ratios of strong nebular emission lines into physical properties such as the gas-phase metallicity, ionization parameter, and ionizing source classification. Most of these lines are generally very bright and thus they can be easily observed even in distant galaxies. Various studies \citep[e.g.,][]{mccall1985, mcgaugh1991, zaritsky1994, kewley2002, Pettini, Pilyugin2005, Maiolino2008, dopita2016, Curti} have used direct auroral metallicity measurements of local galaxies to calibrate strong line ratios and develop empirical and theoretical relations between the strong line ratios and the H\scalebox{0.9}{ II} region metallicity. Table 1 of the recent review by \citet{Maiolino2019} nicely summarizes these conversions. 

Apart from this, ionization sensitive line ratios like [O\scalebox{0.9}{ III}]{$\lambda$4959,5007}/[O\scalebox{0.9}{ II}]{$\lambda$3726,3729} have also been calibrated using photoionization models to measure the ionization state of the gas via the average ionization parameter \citep[e.g.,][]{hainline2009, erb2010, wuyts2012, nakajima2013}. Nebular line ratios have even been extensively used to classify galaxies based on the dominant ionizing source. The most commonly used nebular line diagnostics for this purpose was developed by Baldwin, Phillips \& Terlevich in the early 1980s, and are commonly known as BPT diagrams \citep{baldwin1981, veilleux1987}. The N2-BPT diagram utilizing the line ratios of [O\scalebox{0.9}{ III}]{$\lambda$5007}/H$\beta$ (hereafter O3) vs.\ [N\scalebox{0.9}{ II}]{$\lambda$6584}/H$\alpha$ (hereafter N2) are regularly used as a tool to classify galaxies based on the dominant sources of ionizing radiation into categories such as star-forming galaxies, active galactic nuclei (AGN) hosts, and galaxies with low-ionization emission-line regions (LIERs) (see the recent review by \citealt{kewley2019} for more information). 

Traditionally, the N2-BPT diagram has been calibrated using observations of local galaxies \citep[e.g.,][]{Kewley2001,kauffmann2003}.  That said, there is a growing body of evidence that shows that galaxies at high redshifts do not lie on the same loci of nebular line ratios as local galaxies, therefore complicating the potential conversion of these line ratios to galaxy physical properties at high redshifts. For example, results from the DEEP2 Galaxy Redshift Survey \citep{Shapley_2005, liu2008}, the Keck Baryonic Structure Survey (KBSS) \citep{steidel2014, strom2017}, the MOSFIRE Deep Evolution Field Survey (MOSDEF) \citep{shapley2015, sanders2016}, and the FMOS-COSMOS survey \citep{Kashino_2017}, as well as other observational studies \citep{hainline2009, bian2010, masters2014}, have demonstrated that galaxies at high redshift ($z\sim2$) do not follow the same curve on the BPT diagram as the local star-forming galaxies.  

\begin{figure}[htp]
	\includegraphics[width=\columnwidth]{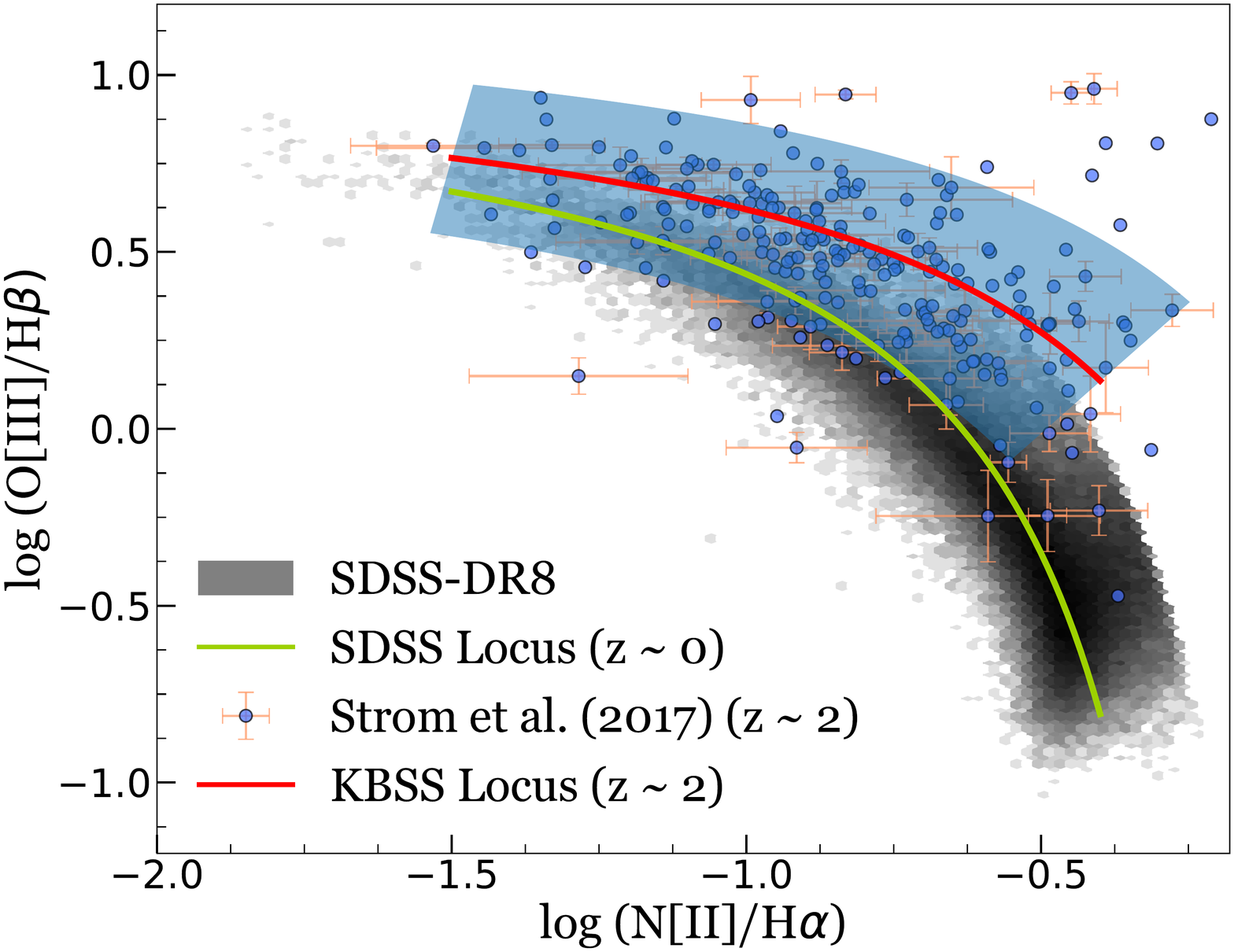}
    \caption{The observed N2-BPT offset between local and high-redshift galaxies. Local galaxies from SDSS-DR8 are shown as black hex bins. Galaxies from KBSS at z $\sim$ 2.3 \citep{strom2017} are shown as blue points. We approximate the galaxy distribution from SDSS-DR8 and KBSS by polynomial fits adopted from \citet{kewley2013a} and \citet{strom2017} , defined in Equation \ref{eq:locus_eqn1} (yellow line) and Equation \ref{eq:locus_eqn2} (red line) respectively. The blue shaded region represents the intrinsic scatter of $\pm$0.18 dex relative to the best-fit curve (red line).}
    \label{fig:offset}
\end{figure}

In Figure \ref{fig:offset}, we summarize these trends. We show the observed N2-BPT offset at high redshifts. The Sloan Digital Sky Survey Data Release 8 (SDSS-DR8) galaxies are represented as black hex bins \footnote{In the figures throughout the paper we only plot the SDSS galaxies that are classified as star forming in the SDSS-DR8 catalog. For more info on how the galaxies were classified see \citet{Brinchmann2004}.} and the galaxies from KBSS at $z\sim2.3$ are shown as blue points. The galaxy distribution in both cases is approximated by the functional form O3 = a/(N2+ b) + c. The values for a, b and c were taken from fits by \citet{kewley2013a} and \citet{strom2017} for the SDSS-DR8 (SDSS locus) and KBSS points (KBSS locus), respectively. In Equations \ref{eq:locus_eqn1} and \ref{eq:locus_eqn2} we show their full functional form. Quantitatively, the N2-BPT offset can be described on average as either $\Delta$ O3 = 0.26 dex or $\Delta$ N2 = 0.37 dex \citep{strom2017}.
\newline
\newline
SDSS locus (z $\sim$ 0):
\begin{equation}\label{eq:locus_eqn1}
O3 = {\Big(\frac{0.61}{N2 + 0.08}\Big)} + 1.1    
\end{equation}
\newline
\newline
KBSS locus (z $\sim$ 2):
\begin{equation}\label{eq:locus_eqn2}
O3 = {\Big(\frac{0.61}{N2 - 0.22}\Big)} + 1.12    
\end{equation} 
\newline

What physical processes drive this offset?  Studies that span a diverse range of techniques, including spectral modeling, photoionization modeling, and hydrodynamic simulations, offer a wide range of possibilities. These include higher ionization parameters \citep[e.g.][]{brinchmann2008, kewley2013a, Kashino_2017, hirschmann2017}, increased contributions from diffuse ionized gas, and varying abundance ratios \citep[e.g.][]{masters2014, shapley2015, Jones2015, sanders2016, cowie2016}. More recently, a number of observational studies \citep{steidel2016, strom2017, strom2018, Shapley_2019, Sanders_direct2020, Topping2020a, Topping2020b, Runco2021} have concluded that the offset in N2-BPT space may be attributed to a harder ionizing spectrum at fixed nebular metallicity, due primarily to stellar  $\alpha$-enhancement.

In this study, we try to understand the origin of this offset by self-consistently modeling a large sample of galaxies at different redshifts using hydrodynamic simulations and photoionization models. We conduct constrained numerical experiments and assess the impact of varying ionization parameters by changing the ionizing photon flux and H\scalebox{0.9}{ II} region densities, increasing radiation field hardness by including the effects of $\alpha$-enhancement and stellar rotation, varying chemical abundances, and stellar physics on the location of galaxies in BPT space.

This paper is organized as follows: In \secref{sec:method} we describe our numerical methodology. In \secref{sec:bpt_results} we present our results at low and high redshifts. In \secref{sec:possible explaination} we test different possible drivers for the subtle mismatch between simulations and observations. In \secref{sec:discussion} we discuss our results in the context of other similar studies and go into some caveats of our model and in \secref{sec:summary} we give a summary of our main conclusions.

\section{Modeling Nebular Emission from Galaxy Formation Simulations}\label{sec:method}
Modeling nebular line emission from H\scalebox{0.9}{ II} regions in galaxy simulations requires modeling a diverse range of physical processes that we summarize here. We first simulate galaxies in a hydrodynamic cosmological simulation. In post-processing, we run \textsc{cloudy} photoionization models on all the young stars. This radiation emergent from these H\scalebox{0.9}{ II} regions is then attenuated as it traverses the dusty, diffuse interstellar medium (ISM) and escapes the galaxy. In what follows, we expand upon this methodology in greater detail.

\subsection{\textsc{simba} simulations}
In this study, we make use of snapshots from the \textsc{simba}  hydrodynamic simulation \citep{simba}. \textsc{simba} is an updated version of the \textsc{mufasa} hydrodynamic simulation code \citep{dave2016mufasa}. It uses \textsc{gizmo}'s \citep{hopkins2015} meshless finite mass (MFM) technique to model hydrodynamics. It includes a H$_2$ based star formation calculated using sub-grid models from \citet{krumholz2009}. The radiative cooling from metals and photoionization heating are handled using \textsc{grackle-3.1} library \citep{smith2017grackle}. The stellar feedback is modeled as a two-phased decoupled kinetic outflow with 30\% hot component \citep{dave2016mufasa}. \textsc{simba} also includes black hole growth via two accretion modes: torque-driven cold accretion \citep{angles2017a} and Bondi accretion \citep{bondi1944}, but only from the hot halo. These simulations reproduce many observed galaxy scaling relations including the galaxy stellar mass function, star-forming main sequence, gas-phase, and stellar mass-metallicity relation both at low and high redshifts. 

An important feature of \textsc{simba} is that it includes on-the-fly dust production and destruction. Dust production takes place via Type II supernova and AGB stars, whereas dust can be destroyed by thermal sputtering, shocks, or can be consumed during star formation. These models have demonstrated success in matching observed dust-to-gas and dust-to-metals ratios in galaxies near and far \citep{li2019}, as well as high-redshift submillimeter galaxy abundances \citep{lovell2021}. Having an accurate model of dust production and destruction is important in modeling interstellar abundances, as some metals are locked up in dust and are therefore not available for ionization.

We take snapshots from \textsc{simba}, run them through a modified version of \textsc{caesar}\footnote{\url{https://github.com/dnarayanan/caesar}} galaxy catalog generator \citep{Thompson2014}. We filter out all the galaxies that do not have at least one star particle younger than 10 Myr. This selection criterion ensures that the galaxy has young stars, the main contributor to the ionizing flux in our model. For all the analyses in this project, we use a simulation box size of side length  $100 h^{-1}$ Mpc with $1024^3$ particles with a baryon resolution of the order of $10^7$ M$_{\odot}$. We assume a $\Lambda$ cold dark matter cosmology with $\Omega_m = 0.3$, $\Omega_{\Lambda} = 0.7$, and $H_0 = 68$ kms$^{-1}$ Mpc$^{-1}$ \citep{ade2016planck}.

\subsection{Photoionization modeling}
We use the Flexible Stellar Population Synthesis (\textsc{fsps})\footnote{\url{https://github.com/cconroy20/fsps}} \citep{conroy2009,conroy2010}, a stellar population synthesis code, to assign a simple stellar population (SSP) to all the star particles in every model galaxy. We examine the impact of SPS model choice later on in the paper in \secref{sec:possible explaination}. \textsc{fsps} comes with a built-in option to include nebular emission. It is implemented using prepackaged \textsc{cloudy} lookup tables which were generated using \textsc{cloudyfsps} \footnote{\url{http://nell-byler.github.io/cloudyfsps/}} \citep[For a detailed description refer to][]{byler2017}. That said, these precomputed lookup tables are only available for a single fixed value for physical properties like hydrogen density, initial mass function, abundance ratios, and spectral libraries. Because a major goal of our project is to understand the impact of various physical effects on the observed location of our model galaxies in BPT space, we run \textsc{fsps}  with nebular line emission turned off and instead compute nebular emission by running \textsc{cloudy} for every individual star particle in each of our model galaxies. Our implementation makes use of \textsc{cloudy} version 17.00 \citep{ferland2017} and borrows heavily from the methodology used in \textsc{cloudyfsps}.

\subsubsection{Modeling parameters}\label{sec:model_param}
Previous theoretical studies in this area typically model each galaxy as a single H\scalebox{0.9}{ II} region, with one set of physical properties \citep[such as their ionization parameter, metallicity, and density; e.g.][]{kewley2013a}. In our model, we consider galaxies to be comprised of ensembles of H\scalebox{0.9}{ II} regions spanning a range of physical properties, where every H\scalebox{0.9}{ II} region surrounding young stars can contribute to the integrated nebular line flux. In what follows, we describe how we model the emission from these H\scalebox{0.9}{ II} regions.  

\paragraph{Ionizing photon production rate}
For calculating nebular emission, we are primarily interested in photons that have energy greater than $13.6$ eV, the ionization potential of hydrogen. Thus, the number of ionizing photons emitted by the source per second is defined as 

\begin{equation}\label{q_eqn}
Q = M_*\int_{\nu{_0}}^{\infty}{\frac{L_\nu}{h\nu} d_\nu}    
\end{equation}
where $L_\nu$ is taken directly from the spectral energy distribution (SED) generated by \textsc{fsps} with nebular emission turned off. 

\paragraph{Assumed geometry}
While H\scalebox{0.9}{ II} regions surrounding massive stars likely have complex geometries owing to stellar winds and magnetic fields \citep[e.g.][]{osterbrook2006,pellegrini2007, ferland2008}, we necessarily must consider relatively simplified geometries owing to the sub-resolution nature of our modeling. In \textsc{cloudy}, the H\scalebox{0.9}{ II} region geometry can be set as either spherical ($\Delta R/R_{\rm inner} \geqslant 3$), a thick shell ($\Delta R/R_{\rm inner} < 3$) or plane parallel ($\Delta R/R_{\rm inner} < 0.1$). Here $\Delta R$ is the thickness of the H\scalebox{0.9}{ II} region and $R_{\rm inner}$ is the distance between the star and the illuminated face of the cloud. For this study, we assume the H\scalebox{0.9}{ II} region geometry to be spherical and fix the $R_{\rm inner}$ at 10$^{17}$ cm. This value for $R_{\rm inner}$ was chosen so that it is small enough that the geometry stays spherical for all the cases when we vary different H\scalebox{0.9}{ II} region properties in \secref{sec:possible explaination}.  In \secref{sec:rinner} we explore the impact of geometry on our model results. We find that varying H\scalebox{0.9}{ II} region geometry leads to an uncertainty of about 0.2 dex in N2 and 0.1 dex in O3.

\begin{figure}[htp]
	\includegraphics[width=\columnwidth]{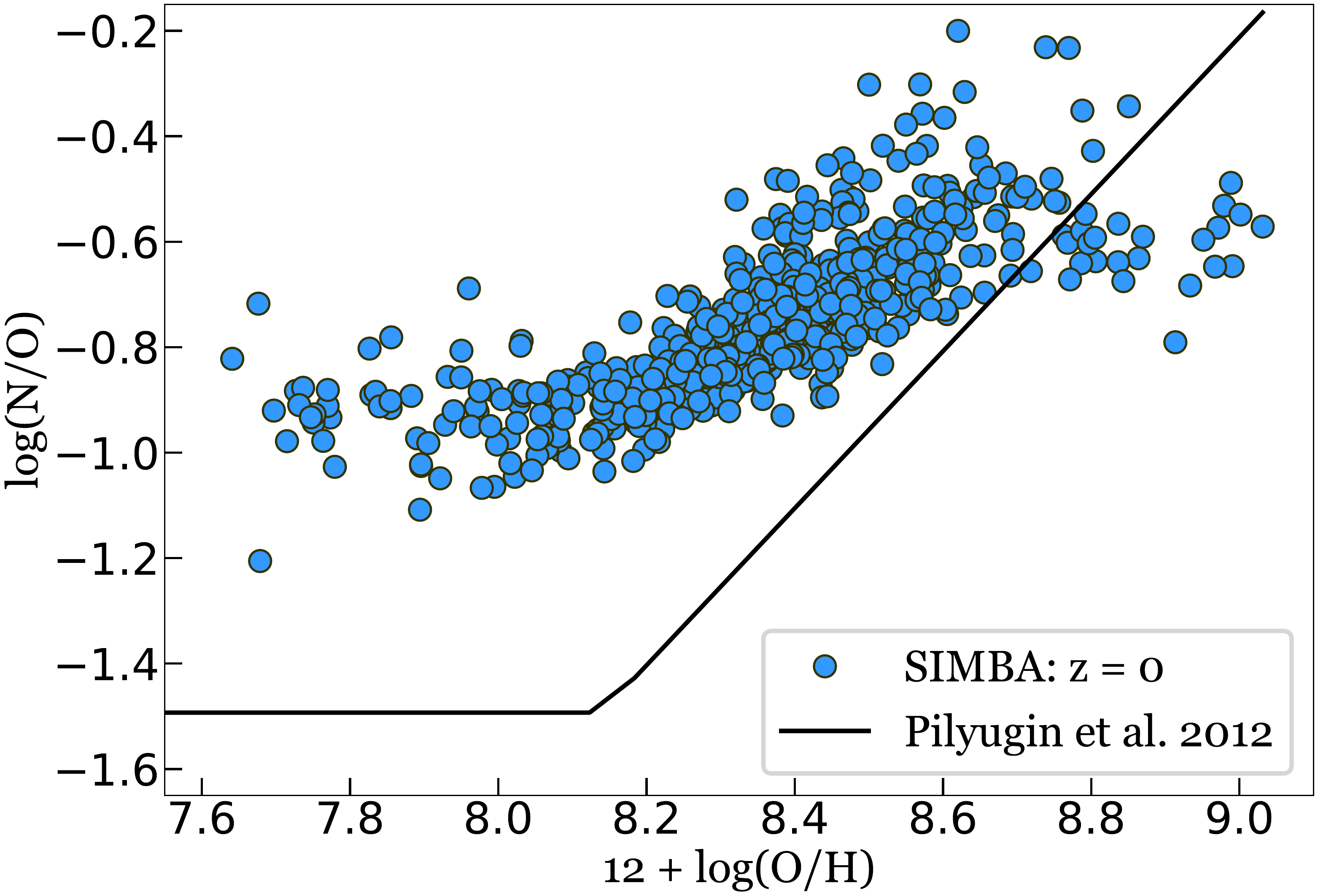}
    \caption{Log (N/O) vs. log(O/H) relation for \textsc{simba} $z=0$ galaxies (8.4 $<$ log($M_{*}/M_\odot$) $<$ 12.6, shown in blue). The observed relation from \citet{pilyugin2012} (Equation \ref{pilyugin_eqn}) is plotted as the black line.}
    \label{fig:NO_z0}
\end{figure}

\paragraph{Stellar and gas properties}\label{gas_properties}
While the time stepping in the hydrodynamic simulation is typically much less than 1 Myr, the minimum age of a star particle in the simulations is set to a floor of 1 Myr, in order to satisfy the constraints within {\sc fsps}. We model the H\scalebox{0.9}{ II} region as a static dust-free sphere with a constant density. We initially assume a Chabrier initial mass function (IMF) and a constant H\scalebox{0.9}{ II} region hydrogen density (n$_H$) of 100 cm$^{-3}$, but vary these assumptions in the analysis that follows. The gas metallicity and age are assumed to be the same as that of the parent star particle. \textsc{simba} tracks the production of 10 elements apart from hydrogen (He, C, N, O, Ne, Mg, Si, S, Ca, and Fe) for each star particle. In general, we use the abundances reported by \textsc{simba} as input to our \textsc{cloudy} model for all the elements. That said, the {\sc simba} model does a relatively poor job reproducing the observed log (N/O) versus log(O/H) ratios for $z=0$ galaxies. We demonstrate this in  Figure \ref{fig:NO_z0}, where we take the SFR-weighted average nitrogen and oxygen abundances and plot the log(N/O) versus log(O/H) relation for \textsc{simba} $z=0$ galaxies (blue points) along with the observed relation (black line) from SDSS defined by Equation \ref{pilyugin_eqn} \citep{pilyugin2012}.
\begin{equation}\label{pilyugin_eqn}
\begin{aligned}
\mathrm{log (N/O)} &= -1.493 \\
&\text{ for } 12 + \mathrm{log(O/H)} < 8.14, \\
&= 1.489 \times [12 + \mathrm{log(O/H)}] - 13.613\\
&\text{ for } 12 + \mathrm{log(O/H)} > 8.14
\end{aligned}
\end{equation}
As can be seen from Figure \ref{fig:NO_z0} the \textsc{simba} $z = 0$ galaxies (8.4 $<$ log($M_{*}/M_\odot$) $<$ 12.6) do not fall on the observed relation, especially at lower metallicity. Thus, instead of taking nitrogen abundances from the simulation, we set nitrogen abundance according to Equation \ref{pilyugin_eqn}. 

\paragraph{Spectral library and stellar isochrones}
For this study, unless otherwise mentioned, we make use of Medium-resolution Isaac Newton Telescope library of empirical spectra (MILES) spectral library \citep{sanchez2006, falcon2011}. MILES is a medium-resolution (2.3 \AA{} FWHM) spectral library consisting of 985 stars. These stars span a wide range of stellar atmospheric parameters: -2.93 $<$ [Fe/H] $<$ 1.65 dex, 2748 $<$ $T_{\rm eff}$ $<$ 36,000 K, and 0.00 $<$ log $g$ $<$ 5.50 dex . For stellar isochrones we make use of 2007 Padova isochrones \citep{bertelli1994, girardi2000, marigo2008} for stars with $M <$ 70 $M_{\odot}$ and Geneva isochrones \citep{lejeune2001} for stars with $M >$ 70 $M_{\odot}$, with high mass-loss rate tracks taken from \citet{schaller1992} and \citet{meynet2000}.  

\paragraph{Cluster mass}\label{cluster_mass}
The typical star particle masses in cosmological simulations are substantially larger when compared to observed star clusters. This leads to the overproduction of ionizing photons per unit H\scalebox{0.9}{ II} region. A single star particle in our simulation can therefore be considered as a collection of unresolved H\scalebox{0.9}{ II} regions. To accurately capture the sub-resolution physics we model the simulation star particle as being an collection of smaller star particles, whose age and metallicity are same as the parent star particle but their masses follow a power-law distribution, defined as

\begin{equation}\label{eq:cmdf_eqn}
\frac{dN}{dM} \propto M^{\beta}
\end{equation}
where $M$ is the stellar mass. There is no consensus in the literature as to what the exact value of $\beta$ should be. For star clusters younger than 10 Myr observational constraints range from $-1.7 < \beta < -2.2$  \citep{chandar2014, chandar2016, linden2017,larson2020}. For this study, we assume $\beta=-2.0$. We subdivide each star particle into 6 mass bins for which our results are converged (see Appendix \ref{sec:massbin}).

We use \textsc{powderday}\footnote{\url{https://powderday.readthedocs.io/en/latest/}} \citep{powderday}, a dust radiative transfer package as a high-level wrapper to do all of our photoionization calculations, and all of the aforementioned physical modules have been implemented into this publicly available code. An additional benefit of working within the {\sc powderday} framework is that the emission from star particles and H\scalebox{0.9}{ II} regions are attenuated by the diffuse dust in the ISM of each galaxy. In detail, the first step is to divide all the star particles younger than 10 Myr into a collection of smaller star particles all of whom have the same age and metallicity as the parent star particle, but their masses follow the cluster mass distribution. This distribution is broken up in such a way that the total mass of the smaller star particles sum to the total mass of the parent star particle. Then, in every galaxy, all the star particles are assigned an SSP using \textsc{fsps} via its python wrapper \textsc{python-fsps}\footnote{\url{http://dfm.io/python-fsps/current/}} \citep{python_fsps}. For all stellar particles younger than 10 Myr, we run \textsc{cloudy} using the model parameters defined above and add the resultant nebular line fluxes and continuum to the SSP continuum to derive the resultant output SED with nebular emission. Once all the particles are populated, we use \textsc{hyperion}\footnote{\url{http://www.hyperion-rt.org/}} \citep{Hyperion} to perform the dust radiative transfer and compute the SED along a given line of sight. The required line fluxes are then extracted from the SED by fitting Gaussian profiles to the emission lines.

\section{Basic Results: Our Model Galaxies on The N2-BPT Diagram}\label{sec:bpt_results}

\subsection{Local galaxies}
\label{section:local_galaxies}
\begin{figure}[htp]
	\includegraphics[width=\columnwidth]{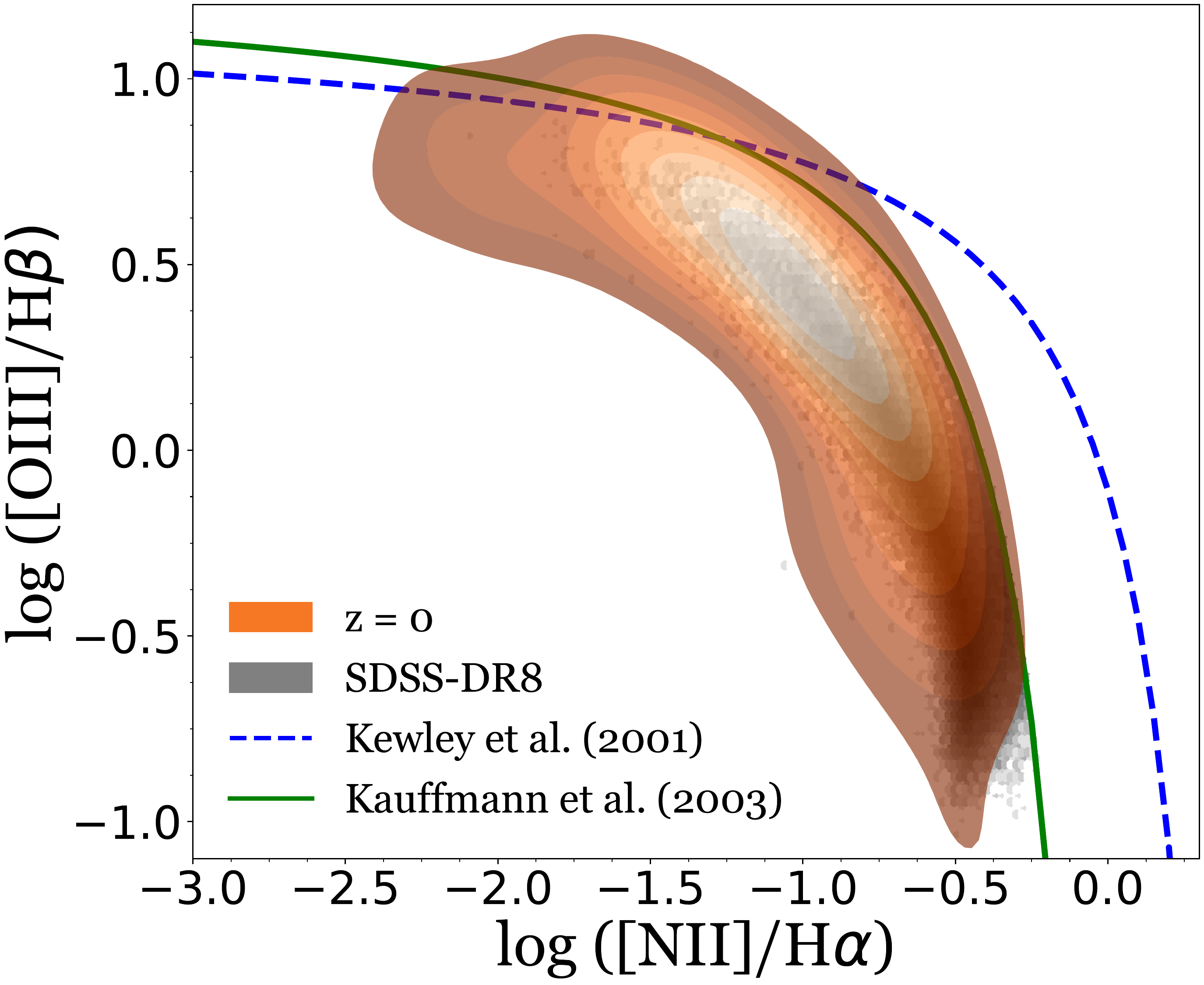}
    \caption{{\sc simba} simulated galaxies at $z=0$ on the N2-BPT diagram (8.4 $<$ log($M_{*}/M_\odot$) $<$ 12.6). Galaxies from \textsc{simba} are shown as orange contours and observations from SDSS-DR8 represented by black hex bins. The division between AGN and starburst galaxies from \citet{Kewley2001} and  \citet{kauffmann2003} is shown by the dashed blue line and green line ,respectively.}
    \label{fig:z0}
\end{figure}

Before attempting to model high-redshift galaxies, we first ascertain that our model reproduces observations of local galaxies. In Figure \ref{fig:z0} we show the location of galaxies from the $z=0$ Mpc \textsc{simba} snapshot (8.4 $<$ log($M_{*}/M_\odot$) $<$ 12.6) on the BPT diagram. The \textsc{simba} galaxies are represented as the orange contours with the observations from SDSS-DR8 \citep{aihara2011} shown as black hex bins. As can be seen from Figure \ref{fig:z0} our model galaxies lie in the same general area on the BPT diagram as the SDSS galaxies, which is encouraging. 

\begin{figure*}
\centering
  \includegraphics[width=\textwidth]{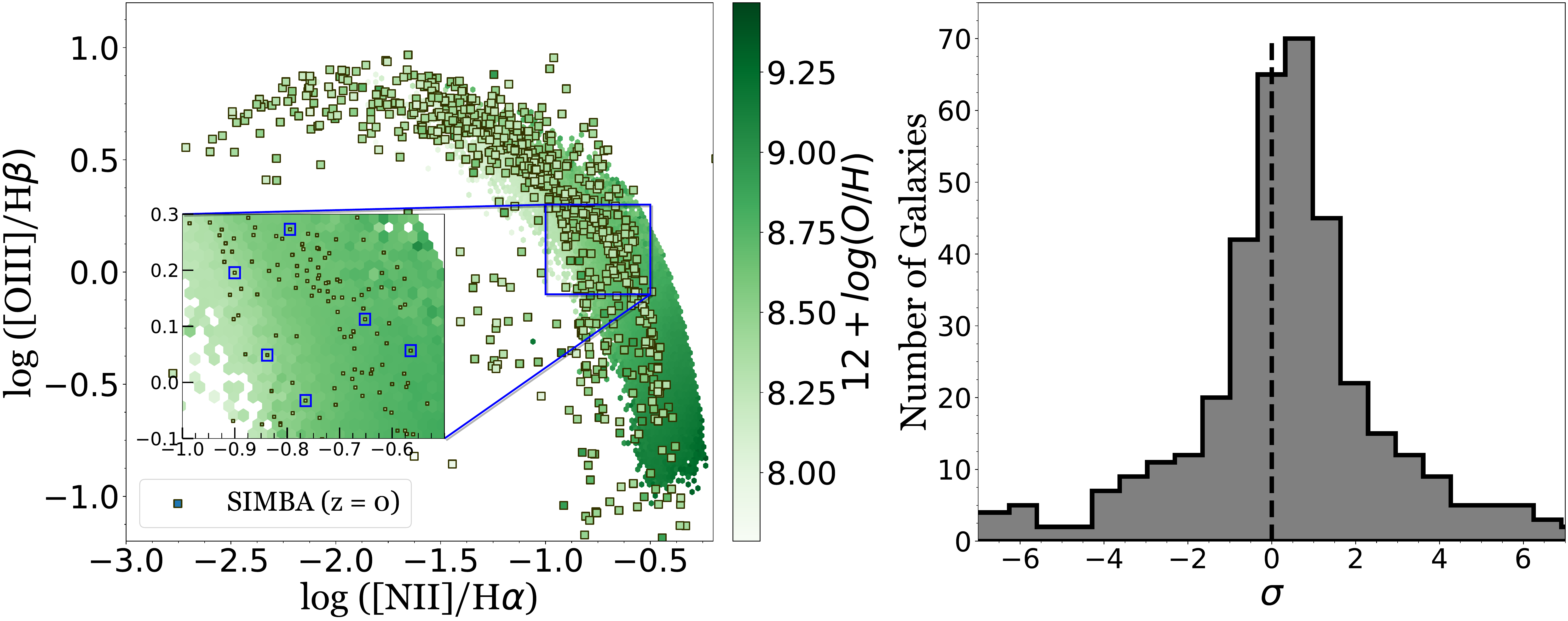}
  \caption{Comparison between $z=0$ simulated galaxies and SDSS galaxies on the BPT diagram.  \textbf{Left panel:} shows the relative positions of \textsc{simba} z = 0 and SDSS galaxies in metallicity space on the N2-BPT diagram. \textsc{simba} and SDSS galaxies are represented as squares points and hex bins respectively. The data is color coded to metallicity and shows a reasonable correspondence. \textbf{Right panel:} shows the distribution of \textsc{simba} z = 0 galaxies with metallicity within a given standard deviation ($\sigma$) of the SDSS galaxies in a 0.02 dex square box. Overall, our model galaxies have comparable metallicities to those observed in the SDSS at a fixed location in N2-BPT space. See text in \secref{section:local_galaxies} for details.}
  \label{fig:z0_metal}
\end{figure*}

To further test our model at $z=0$, we compare the metallicity of our model galaxies to the metallicity (defined as 12 + log(O/H)) of the observed SDSS galaxies in their vicinity on the N2-BPT diagram. The location of a galaxy on the N2-BPT diagram is highly sensitive to metallicity. Decreasing the metallicity lowers the nitrogen abundance due to secondary nitrogen production, which decreases the [N\scalebox{0.9}{ II}] flux. At the same time, the dependence of [O\scalebox{0.9}{ III}] on metallicity is a bit more complicated. Apart from changing the oxygen abundance, varying the metallicity also modifies the collisional excitation rates of O++. At high metallicity, collisional excitation is not favored due to low temperatures. Decreasing the metallicity increases the equilibrium temperature, which leads to an increase in the [O\scalebox{0.9}{ III}] flux. As the metallicity continues to decrease, there comes to a point where the decrease in abundance starts to balance the increasing collisional rates. This is where we see a peak
in the O3 line ratio and a subsequent decrease in metallicity decreases the O3 line ratio (see Figure \ref{fig:z0_metal}). Metallicity also has a second-order dependence through the ionization parameter. Stars with lower metallicity (iron-poor stars) produce more ionizing photons per unit stellar mass and therefore drive a higher ionization parameter at fixed gas geometry and density.

In Figure \ref{fig:z0_metal} (left panel) we show the SDSS and the \textsc{simba} $z = 0$ galaxies color coded based on their metallicity. To quantify where our galaxies lie in the metallicity space as compared to the SDSS galaxies we take all SDSS galaxies that lie within a square box of 0.01 dex around every \textsc{simba} galaxy (see an example in the inset in the left panel of Figure \ref{fig:z0_metal}) and then calculate within how many standard deviations from the mean does the \textsc{simba} galaxy lie. We only consider those \textsc{simba} galaxies that have at least 10 SDSS galaxies within the box. The distribution is plotted in Figure \ref{fig:z0_metal} (right panel) and we find that out of the \textsc{simba} galaxies that meet the above criteria about 75\% and 45\% of them lie within 3$\sigma$ and 1$\sigma$ respectively. 

The distribution in Figure \ref{fig:z0_metal} (right panel) demonstrates a reasonable correspondence between observations and simulations. That said, there is an offset in the sense that the \textsc{simba} galaxies have a higher metallicity than SDSS galaxies at the same location on the BPT diagram. This potentially arises from the difference in how metallicity is being calculated for both cases. For the \textsc{simba} galaxies, the metallicity is calculated by taking the SFR-weighted average metallicity of the gas particles.  We note that this SFR-weighted average metallicity is different from the stellar metallicity used to compute nebular emission, though for stars younger than 10 Myr this distinction does not matter much and the stellar metallicity and SFR-weighted gas metallicity should be in reasonable agreement. In contrast, the abundances of SDSS galaxies were derived using the Charlot and Longhetti models \citep{tremonti2004}, which have an uncertainty of a factor of 2. Thus, owing to the difference in how the metallicity is being calculated in the two cases and the inherent uncertainties involved we consider our models a reasonable match for $z=0$ observations and now proceed to understand the physics driving the BPT diagram at $z=2$.

\subsection{The BPT curve at \texorpdfstring{$z \sim 2$}{z~2}}
\begin{figure}[htp]
\centering
  \includegraphics[width=\columnwidth]{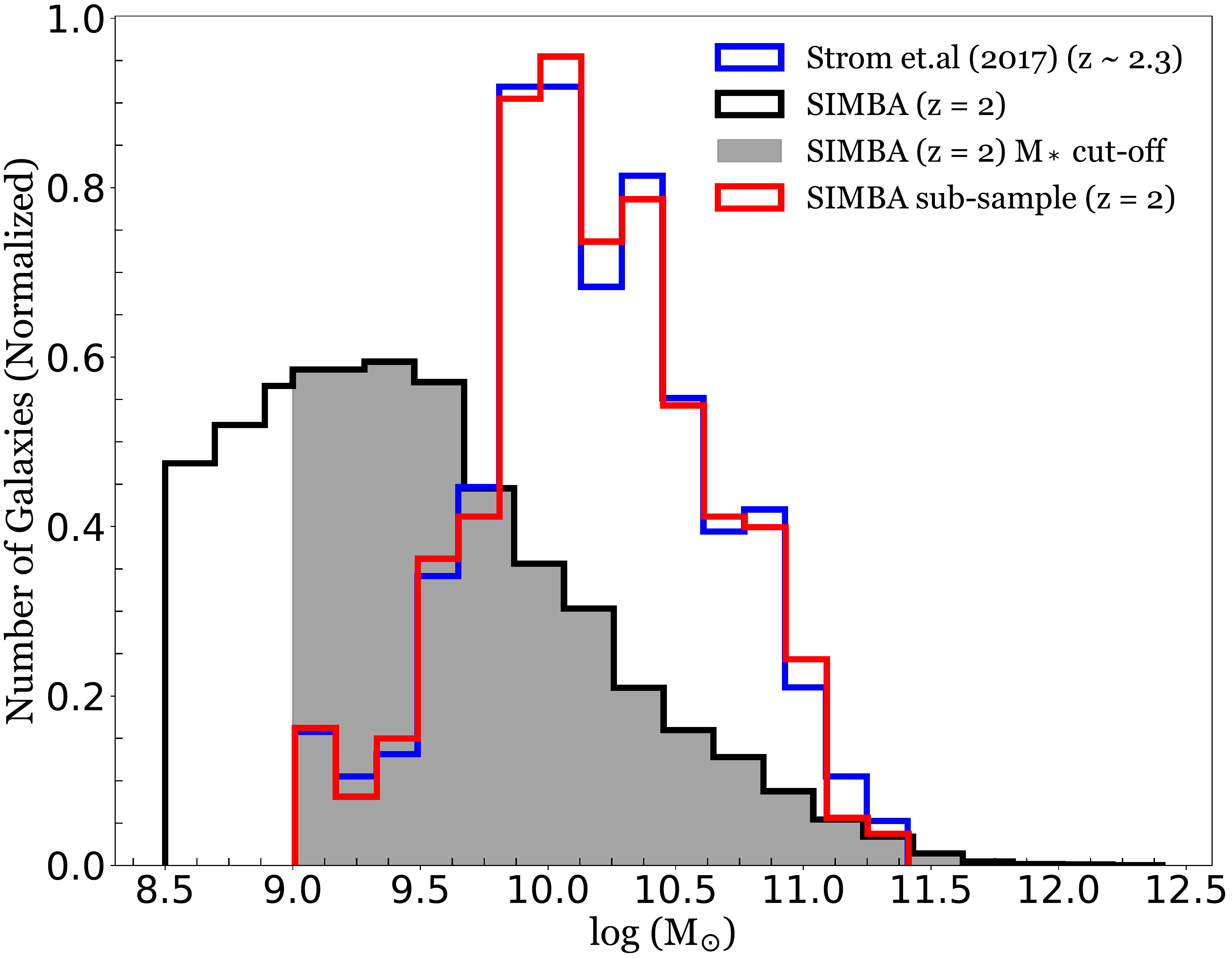}
  \caption{Stellar mass distribution of the {\sc simba} $z=2$ (black) and KBSS galaxies (blue). To make sure the  {\sc simba} $z=2$ galaxies represents the observed KBSS sample we apply a cutoff of $10^9 M_{\odot}$, shown by the shaded region. We further generate a subsample of 1000 galaxies (show in red ) that follow the mass distribution of KBSS galaxies}
  \label{fig:z2_mass}
\end{figure}

Our first step is to examine our default model at $z=2$ in comparison to observations. To do this, we examine the stellar mass distribution of {\sc simba} $z=2$ and the KBSS galaxies, plotted in Figure \ref{fig:z2_mass} as black and blue histograms, respectively. We first apply a stellar mass cutoff of $10^9 M_{\odot}$ in our model galaxies to mimic the rough mass cutoff in the KBSS galaxy sample.  Though the mass ranges for {\sc simba} $z=2$ and the KBSS galaxies are now similar the shape of the two distributions is still quite different: the {\sc simba} $z=2$ snapshot contains many more low mass galaxies than the observed KBSS sample.  To match the KBSS sample, we randomly draw a subsample of $1000$ galaxies from our parent sample of galaxies, but following the KBSS mass distribution.  The resulting mass distribution of our sample of galaxies is shown in red in Figure~\ref{fig:z2_mass}.

\begin{figure*}
	\includegraphics[width=\textwidth]{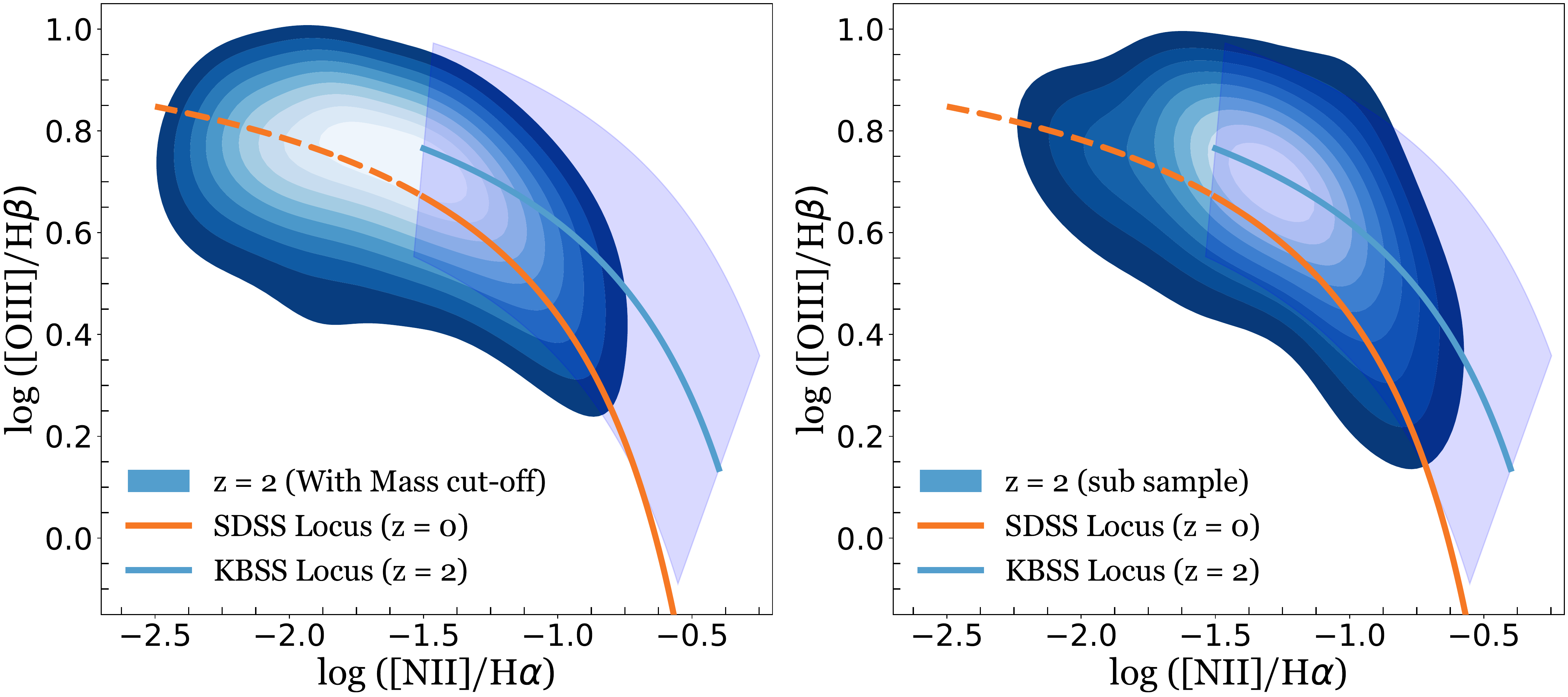}
    \caption{\textsc{simba} $z = 2$ galaxy distribution before (left panel) and after (right panel) performing the subsampling to match the KBSS mass distribution. As can be seen, if we just include mostly high mass galaxies as is the case with the subsample shown on the right panel then the distribution moves toward the observed N2-BPT offset. Thus, we argue that the high-redshift BPT offset naturally arises due to sample selection effects. \textbf{Left panel:} N2-BPT diagram of model galaxies at $z=2$ with a mass cutoff of $10^9 M_{\odot}$ (shaded region in Figure \ref{fig:z2_mass}). \textbf{Right panel:}  N2-BPT diagram of subsample of $z=2$ model galaxies with stellar mass distribution following the observed KBSS mass distribution (red and blue histogram in Figure \ref{fig:z2_mass}). The galaxy distribution of the subsample (right panel) moves toward lower O3 and higher N2 due to the distribution no longer being skewed toward low mass galaxies, giving a better match to the observed KBSS sample.}
    \label{fig:z2}
\end{figure*}

In Figure \ref{fig:z2} (left panel) we plot the N2-BPT diagram of the mass-cutoff sample, showing all simulated galaxies with $\textrm{log(M}_*) >$  9.0. We find that for the mass-cutoff sample the {\sc simba} galaxies mostly lie along the SDSS locus (orange line) not showing any appreciable offset. In the right panel of Figure~\ref{fig:z2}, we plot the N2-BPT diagram that is mass matched to the KBSS sample. We can see that the distribution as whole moves toward lower O3 and higher N2 with the peak now lying on the KBSS locus just at the edge of the observed sample.  This happens because the KBSS mass distribution matched subsample has higher mass (and therefore higher metallicity) galaxies than our parent simulation sample. Thus, we argue that \textit{the observed N2-BPT offset as compared to $z\sim 0$ galaxies arises primarily due to sample selection effects.} A natural prediction from this model is that deeper surveys (with, e.g., James Webb Space Telescope (JWST)), will reveal N2-BPT diagrams at high-redshift that follow a similar arc as $z\sim 0$ galaxies. Moving forward we consider the KBSS mass-matched subsample of 1000 {\sc simba} $z=2$ galaxies as our default sample at $z=2$. 

\begin{figure*}
	\includegraphics[width=\textwidth]{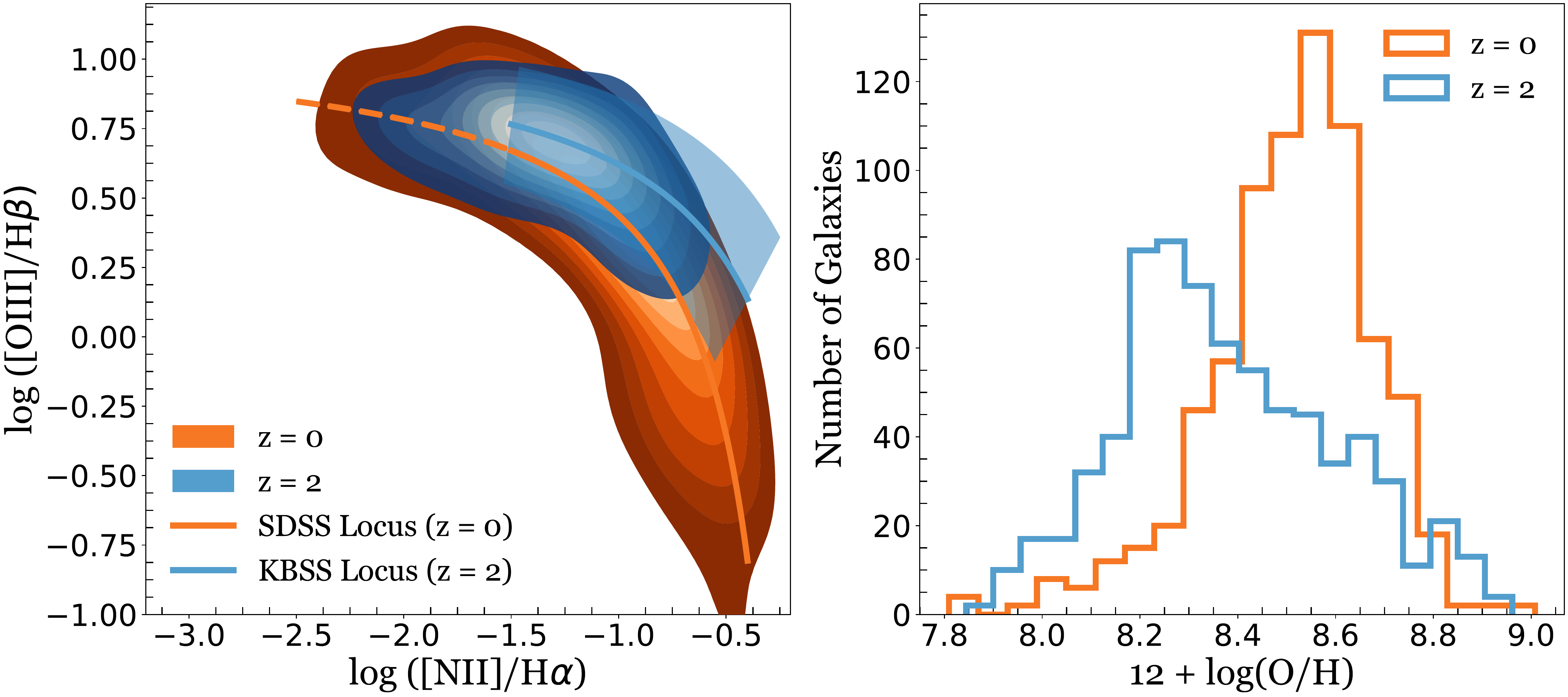}
    \caption{N2-BPT diagram of \textsc{simba} galaxies at $z=0$ and $z=2$. When comparing we can see that the galaxy distribution at $z = 2$ moves higher up on the N2-BPT diagram (see left panel). This occurs due to decreased metallicities at $z=2$. \textbf{Left Panel:} \textsc{simba} simulated galaxies at z = 0 and z = 2 on the N2-BPT diagram, shown as orange and blue density plots respectively. We approximate the galaxy distribution from SDSS and KBSS $z \sim 2$ by polynomial fits adopted from \citet{kewley2013a} and \citet{strom2017}, defined by equation \ref{eq:locus_eqn1} (orange line) and \ref{eq:locus_eqn2} (blue line) respectively. The dashed orange line shows the extrapolation of the SDSS locus beyond the SDSS-DR8 data points. The blue shaded region represent the intrinsic dispersion of $\pm$0.18 dex relative to the best fit curve (blue line). \textbf{Right Panel:} The metallicity distribution of \textsc{simba}  z = 0 and z = 2 galaxies shown by orange and blue histograms respectively.}
    \label{fig:z0Vz2}
\end{figure*}

At the same time, while our default model at $z=2$ has a comparable shape as observed KBSS galaxies, there is a mismatch in the peak of the distribution between simulations and observations: our modeled distribution on average has a higher O3 and lower N2 when compared to the observed KBSS sample.  In Figure \ref{fig:z0Vz2} (left panel), we compare the N2-BPT line ratios of {\sc simba} and $z=0$ and $z=2$ galaxies. Since we have used the same model parameters as the $z=0$ case, we find that the galaxies simply move up along the SDSS locus due to their lower metallicity. This is evident from the metallicity distribution of the {\sc simba} $z=0$ and $z=2$ galaxies, shown in the right panel of Figure \ref{fig:z0Vz2}. This implies that our model parameters may not provide an accurate representation of the H\scalebox{0.9}{ II} regions at high redshift. In the next section, we turn our attention toward investigating the physical impact of our assumptions on the N2-BPT diagram to understand the origin of this subtle mismatch between observations and simulations.

\section{Potential Drivers of mismatch between simulations and observations at \texorpdfstring{High-$z$}{High-z}}\label{sec:possible explaination}

In this section, we perform a series of controlled numerical experiments to understand how different physical effects move galaxies in the N2-BPT diagram, and to attempt to understand the origin of the subtle mismatch between our simulations and observations. We look at the effects of an evolving ionization parameter in \secref{sec:ip_dec}, hydrogen density in \secref{sec:Hden}, abundance ratios in \secref{sec:no_ratio}, and hardening radiation field in \secref{sec:harder_field}. As we will demonstrate, out of these possibilities either an increasing N/O ratio or a decreasing ionization parameter at fixed O/H can move the peak of the galaxy distribution lower along the KBSS locus thus giving a better match to the observations.
\\
\\
\subsection{Ionization parameter}\label{sec:ip_dec}
\begin{figure*}
	\includegraphics[width=\textwidth]{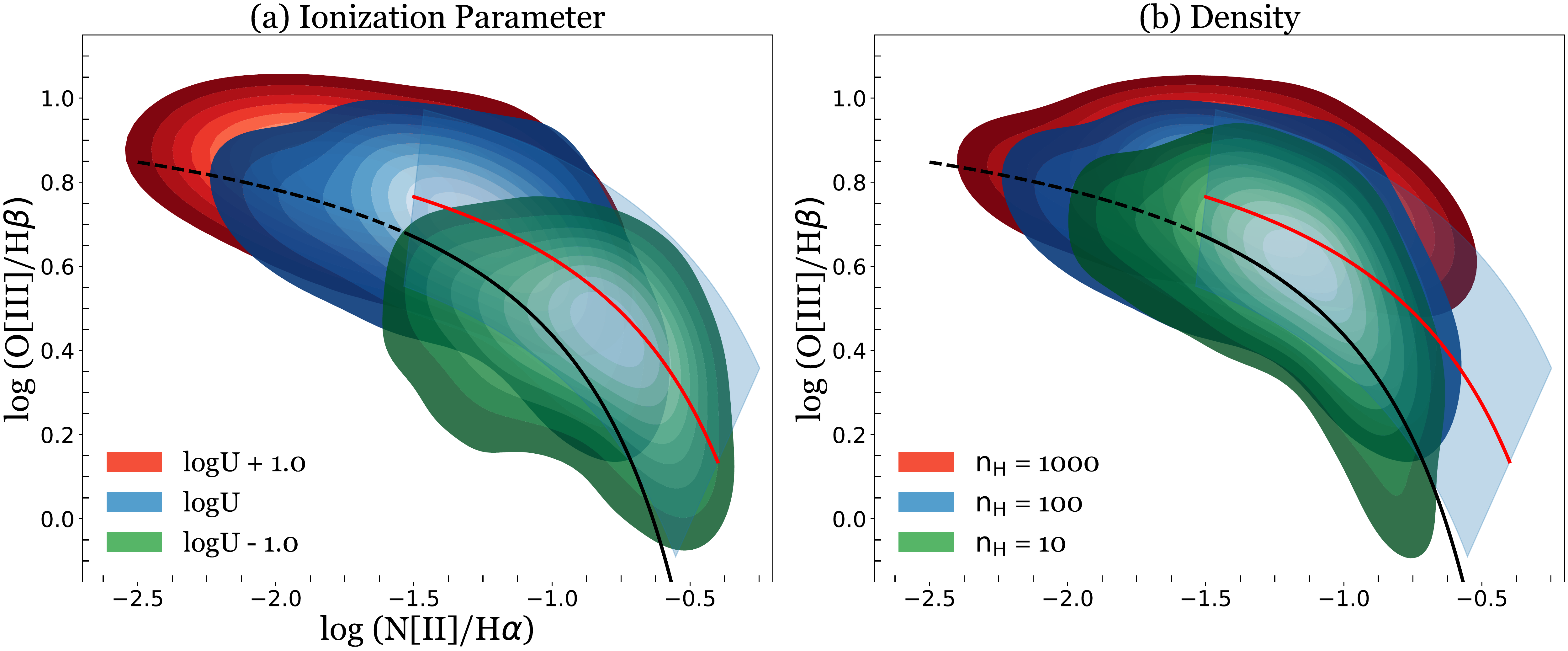}
    \caption{Effects of varying the ionization parameter and density. We find that decreasing the ionization parameter can move the galaxies toward lower O3 and higher N2 better matching the observations (left panel, see the distribution shown in green). \textbf{Left panel:} the \textsc{simba} z = 2 base model shown in blue and models with logU by -1.0 and +1.0 are shown in red and green, respectively. \textbf{Right panel:} $n_{\rm H}$ = 1000, 100 (base model) and 10 shown in red, blue, and green, respectively. In all the figures SDSS and KBSS z $\sim$ 2.3 data points are approximated by polynomial fits adopted from \citet{kewley2013a} and \citet{strom2017}, defined by equation \ref{eq:locus_eqn1} (black line) and \ref{eq:locus_eqn2} (red line), respectively. The dashed black line shows the extrapolation of the SDSS locus beyond the SDSS-DR8 data points. The blue shaded region represent the intrinsic dispersion of $\pm$0.18 dex relative to the best-fit curve (red line)}
    \label{fig:logU}
\end{figure*}
The ionization parameter is the dimensionless ratio between the number of ionizing photons and the number of hydrogen atoms in a medium. It encapsulates the ionization state of an H\scalebox{0.9}{ II} region and along with metallicity is fundamental in determining the position of a galaxy on the BPT diagram. We use the following equivalent definition for the ionization parameter in our analysis:

\begin{equation}\label{eq:U_eqn1}
U = {\Big(\frac{Q}{4\pi\,{R_S}^2\,n{_H}\,c}\Big)}    
\end{equation}

Where, R$_S$ is the Str\"{o}mgren radius given by
\begin{equation}\label{rs_eqn}
R_S = {\Big(\frac{3Q}{4\pi\,n{_H}^2\,\alpha_B}\Big)}^{1/3}    
\end{equation}

On substituting Equation \ref{rs_eqn} back into Equation \ref{eq:U_eqn1} we get
\begin{equation}\label{eq:U_eqn2}
U = {\Big(\frac{Q\,n{_H}\,{\alpha_B}^2}{36\pi\,c^3}\Big)}^{1/3}  \implies  U \propto Q^{1/3} n{_H}^{1/3}
\end{equation}

Here, $Q$ is the number of H-ionizing photons (Equation \ref{q_eqn}), $n_{\rm H}$ is the volume-averaged hydrogen density, $c$ is the speed of light, and $\alpha_{\rm B}$ is the case B recombination coefficient. Note, this definition of ionization parameter is different from what is needed by \textsc{cloudy} as an input. In \textsc{cloudy} you can either provide it the number of ionizing photons per second emitted by the source and the inner radius of the cloud (this is the approach we use) or give it an ionization parameter. Importantly, the ionization parameter that \textsc{cloudy} needs is evaluated at the inner radius ($R_{\rm inner}$). Observationally, a commonly used method of inferring the ionization parameter is by using the line ratio of the same element in different ionization states like [O\scalebox{0.9}{ III}]{$\lambda$5007}/[O\scalebox{0.9}{ II}]{$\lambda$3727}. A number of studies \citep[e.g.][]{Brinchmann2004, liu2008, hainline2009, erb2010, wuyts2012, nakajima2013, shirazi2014, hayashi2015, onodera2016, Kashino_2017} have inferred higher ionization parameters in $z \sim 2$ galaxies as compared to local counterparts with similar mean mass.

Could a varying ionization parameter at fixed O/H drive the galaxies lower along the KBSS locus? To test this we take our base model and manually change the ionization parameter to see how the galaxies move in response on the N2-BPT diagram. It can be seen from Equation \ref{eq:U_eqn2} that the ionization parameter depends on the number of ionizing photons ($Q$) and hydrogen density ($n_{\rm H}$). To change the ionization parameter we keep the $n_{\rm H}$ fixed at the assumed value of $100$ cm$^{-3}$ and change $Q$ appropriately to get the desired value of the ionization parameter. In essence, modulating the number of ionizing photons is equivalent to assuming a different cluster mass distribution slope ($\beta$ in Equation \ref{eq:cmdf_eqn}) or changing the cluster mass distribution function upper and lower-mass limits or even assuming different IMFs. Having more high mass clusters or more massive stars will lead to a greater number of ionizing photons increasing U and vice versa. We would like to emphasize that during this analysis when we vary the ionization parameter the cloud geometry and the gas properties are kept fixed, and the only thing that is changing is the flux of ionizing photons from the stellar source.

In Figure~\ref{fig:logU} (left panel), we show the results of varying the ionization parameter where the base model is shown in blue and the cases with logU increased by 1 dex and decreased by 1 dex are shown in red and green, respectively. As can be seen, decreasing the ionization parameter while keeping all other parameters fixed moves the points toward lower O3 and higher N2. This happens because of the relative populations of double and singly ionized species of an element change with the ionization parameter. Decreasing the ionization parameter means that less oxygen is in O$_{\mathrm{III}}$ versus O$_{\mathrm{II}}$, decreasing the [O\scalebox{0.9}{ III}] flux. At the same time, a lower ionization parameter leads to more of the nitrogen being in the singly ionized state, increasing the [N\scalebox{0.9}{ II}] flux. Thus, in effect, decreasing the ionization parameter moves the BPT curve toward the higher N2 and lower O3 (bottom right), and increasing the ionization parameter moves the curve toward lower N2 and higher O3 (top left). The important point to note is that decreasing the ionization parameter at fixed metallicity moves the peak of galaxy distribution towards lower O3 and higher N2  (bottom right) thus giving a better match to the observed KBSS sample (blue shaded region). In our model the ionization parameter (see equation \ref{eq:U_eqn2}) depends on the ionizing photon production rate (the effects of which we have shown in this section), $R_{\rm inner}$ (this is kept fixed in our modeling), and the hydrogen density. Thus, for completeness, we also investigate varying H\scalebox{0.9}{ II} region densities as a possible origin of changing ionization parameters in galaxies.

\subsubsection{Hydrogen density (\texorpdfstring{n$_H$}{nH})}\label{sec:Hden}
A number of studies have inferred increased ionized-gas densities in high-redshift galaxies as compared to their local counterparts where the inferred value lies within the 10 - 100 cm$^{-3}$ range \citep[e.g.,][]{brinchmann2008, hainline2009, rigby2011, wuyts2012, bian2010, sanders2016, strom2017}. Changes in hydrogen density will impact the ionization state in the \textsc{cloudy} calculations (see Equation \ref{eq:U_eqn2}). An increase in hydrogen density will make collisional de-excitation more probable causing an increase in radiative cooling through optical transitions \citep{gutkin,hirschmann2017}. 

For the base model, we assumed the hydrogen density to be $100$ cm$^{-3}$. To see what effect changing hydrogen density has on the emission-line ratios we reran our model with a hydrogen density of 1000 cm$^{-3}$ and 10 cm$^{-3}$, shown as red and green density plot in Figure \ref{fig:logU} (right panel). We find that increasing the density by an order of magnitude increases O3 by about 0.1 dex on average. As for the N2, it decreases by about 0.1 dex for low metallicity galaxies, whereas it is almost unchanged for high metallicity galaxies. This result is in tension with the conclusions of \cite{sanders2016}, who found that increasing density from 25 cm$^{-3}$-250 cm$^{-3}$ leads to an increase of about 0.1 dex in both N2 and O3. The reason for the disagreement may be attributed to the fact that in the analysis of \citeauthor{sanders2016},  they vary the density while keeping the ionization parameter fixed. As can be seen from Equation \ref{eq:U_eqn2}, to force the ionization parameter to stay constant while increasing the hydrogen density, either the ionizing photon rate from stellar sources has to increase proportionally or the inner radius of the cloud has to decrease proportionally. Therefore, the stellar properties are not kept constant when the density is varied. In our analysis, we keep all the stellar and cloud properties except the hydrogen density constant. The net overall effect of this is that increasing the density leads to a decrease in N2 and an increase in O3. This is similar to what happened when we manually increased the ionization parameter (Figure \ref{fig:logU} (left panel)).

In the end, we find that \textit{decreasing the ionization parameter at fixed O/H can move the peak distribution toward lower O3 and higher N2 providing a better match between the simulations and the observed KBSS sample.}

\subsection{Abundances Patterns (N/O Ratio)}\label{sec:no_ratio}
\begin{figure}[htp]
\centering
  \includegraphics[width=\columnwidth]{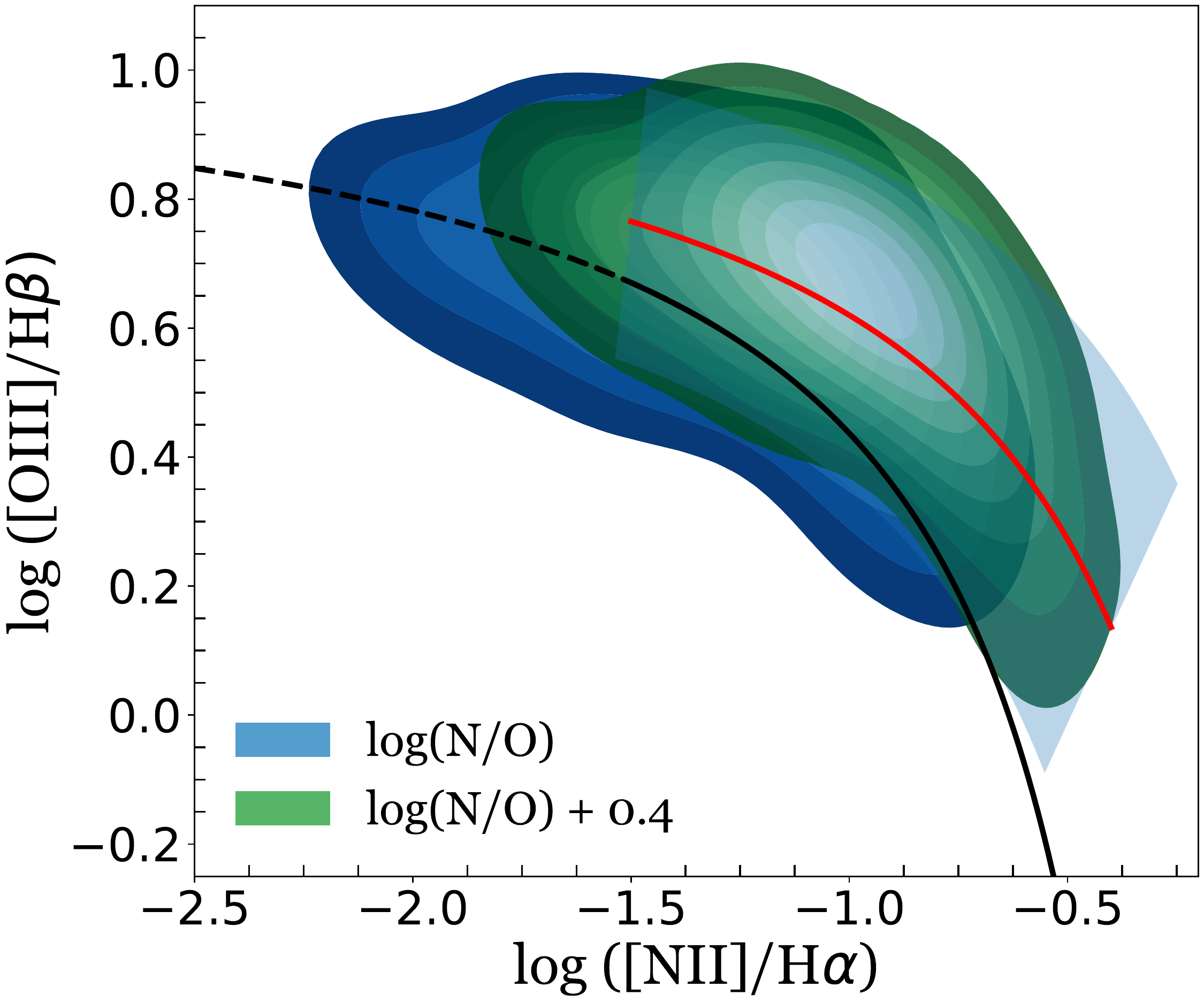}
  \caption{Effects of having a higher N/O ratio. \textsc{simba} z = 2 galaxies with base model and log(N/O) + 0.4 are shown in blue and green, respectively. As can be seen increasing the N/O  ratio leads to an increase in the N2 line ratio and the \textsc{simba} z = 2 galaxy distributions moves toward higher N2 providing a better match to the observed KBSS sample. The SDSS and KBSS z $\sim$ 2.3 data points are approximated by polynomial fits adopted from \citet{kewley2013a} and \citet{strom2017}, defined by equation \ref{eq:locus_eqn1} (black line) and \ref{eq:locus_eqn2} (red line), respectively. The dashed black line shows the extrapolation of the SDSS locus beyond the SDSS-DR8 data points. The blue shaded region represent the intrinsic dispersion of $\pm$0.18 dex relative to the best-fit curve (red line).}
  \label{fig:NO_ratio}
\end{figure}
Having a higher N/O ratio will lead to an increase in [N\scalebox{0.9}{ II}] flux, which moves the galaxies toward the right along the x-axis on the N2-BPT plane, therefore, more closely matching the observed high-redshift galaxy distribution. To test the effects of having a higher N/O ratio we rerun the z = 2 snapshots with a N/O ratio increased by $0.4$ dex. The result is shown in Figure \ref{fig:NO_ratio} and it can be seen that increasing the N/O abundance ratio moves the peak downward along the KBSS locus. \textit{Thus, increasing the N/O ratio at fixed O/H can move the peak of the galaxy distribution toward lower O3 and higher N2 giving a better match to the observations.}

\subsection{Harder Radiation Field}\label{sec:harder_field}
\begin{figure*}
	\includegraphics[width=\textwidth]{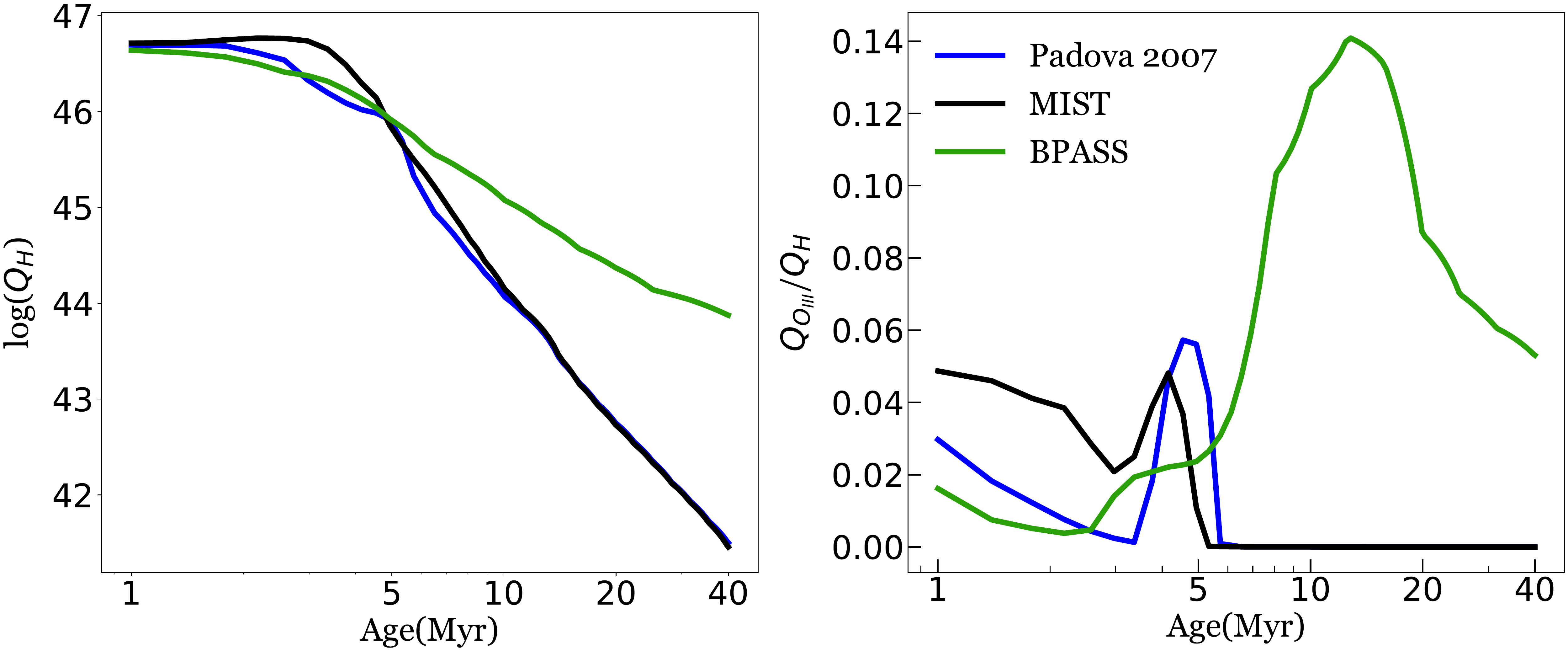}
    \caption{Time evolution of the number of ionizing photons emitted and the radiation field hardness for an SSP of 1 $M_\odot$. \textbf{Left panel:} variation in the number of ionizing photons with age for different models. Models with Padova 2007 isochrones (base model), MIST isochrones (stellar rotation), and BPASS (binary stars) shown in blue, black, and green, respectively. \textbf{Right panel:} hardness of radiation field varying with age. We define the hardness as the ratio of the number of photons that can doubly ionize oxygen ( $Q_\textrm{[O\scalebox{0.9}{ III}]}$: $h\nu > 35$ eV) to the number of H-ionizing photons ($Q_{\rm H}$: $h\nu > 13.6$ eV). Models with Padova 2007 isochrones (base model), MIST isochrones (stellar rotation), and BPASS (binary stars) shown blue, black, and green, respectively.}
    \label{fig:Qh}
\end{figure*}
\begin{figure*}
\centering
	\includegraphics[width=\textwidth]{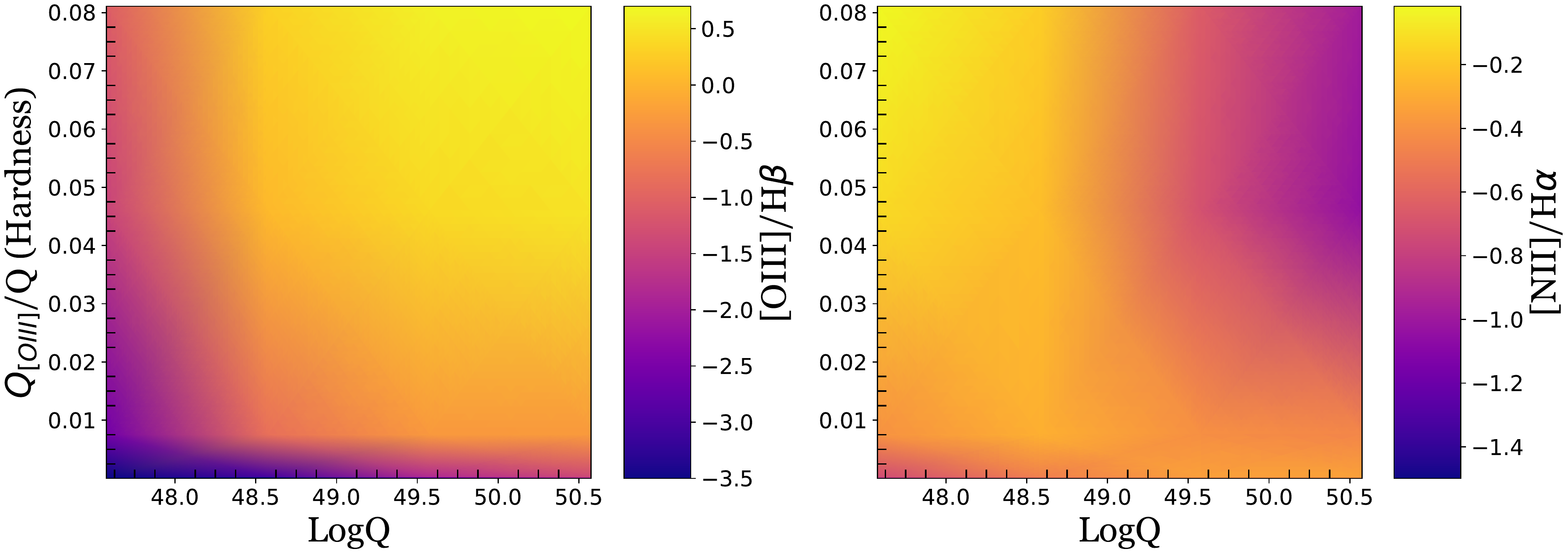}
    \caption{We make use of a 1 $M_\odot$ SSP to test the dependence of the line ratios on the hardness of the stellar spectrum and the number of H-ionizing photons ($Q$). We use the ratio of the number of photons that can doubly ionize oxygen ($Q_\textrm{[OIII]}$) to the number of photons that can ionize hydrogen ($Q$) as a proxy for hardness. We find that when hardness is increased O3 always increases irrespective of $Q$ whereas there is a bimodality in N2. With increasing hardness, N2 decreases for high values of $Q$ and increases for low values of $Q$. \textbf{Left panel:} Variation of the [O\scalebox{0.9}{ III}]/H$\beta$ ratio with changing hardness and log$Q$ of the input spectrum. \textbf{Right panel:} Variation of the [N\scalebox{0.9}{ II}]/H$\alpha$ ratio with changing hardness and log$Q$ of the input spectrum.}
    \label{fig:ae_pcolor}
\end{figure*}

Recent studies like \citet{kewley2013a}, \citet{steidel2014}, \citet{steidel2016}, \citet{strom2017}, \citet{Shapley_2019}, \citet{Sanders_direct2020}, \citet{Topping2020a}, \citet{Topping2020b}, and \citet{Runco2021} have argued that a harder field at fixed metallicity may be the primary driver of the high-redshift BPT offset. Many of these studies argue for $\alpha$-enhancement as a source of this harder radiation field. That said, we find that the observed N2-BPT offset at high redshifts just can be explained by sample selection effects.  In light of this, in this subsection, we investigate whether having a harder radiation field can provide a better match between the peaks of the distributions in the simulations and observations (i.e., whether the radiation field can move our simulated galaxies down the $z=2$ N2-BPT arc). Before trying to understand how the hardening of the radiation field can affect the position of galaxies on the N2-BPT diagram we first look at some of the scenarios through which we can invoke a harder radiation field in our models.

\subsubsection{\texorpdfstring{$\alpha$-enhancement}{a-enhancement}}\label{sec:AE}
Elements synthesized by the $\alpha$-process namely C, O, Ne, Mg, Si, S, Ar, Ca, and Ti are known as $\alpha$-elements. These elements are primarily produced in Type II supernovae (\citep[SNe;][]{woosley1995}. In contrast, the iron-peak elements (Fe, Cr, Mn, Fe, Co, and Ni) are produced mainly through Type Ia SNe \citep{tinsley1979, greggio1983}. The progenitors of Type II SNe are short-lived massive stars ($\leqslant$ 10Myr) whereas Type Ia SNe naturally occurs over longer time scales (100 Myr - 1 Gyr). This temporal delay in the enrichment of iron-peak elements means that $\alpha$/Fe abundance ratios like O/Fe can be used to trace the star-formation history and timescales \citep{trager2000, puzia2005, woodley2010, thomas2010, johansson2012, conroy2013, Hughes2020}.  Some studies \citep{steidel2016, matthee2018, strom2018}  have found that in z $>$ 1 galaxies, massive stars that produce the bulk of the ionizing radiation have an O/Fe $>$ O/Fe$_\odot$ i.e they are $\alpha$-enhanced. This has been attributed to relatively short star formation timescales in these galaxies. The opacity of massive stars is mainly governed by the line blanketing from iron-peak elements. Therefore, being $\alpha$-enhanced or iron-poor makes these stars less opaque to ionizing photons. This leads to higher effective temperatures, which in turn make the radiation field harder \citep{eldridge2017, stanway2018}. Since the current version of \textsc{fsps} does not support non-solar abundance ratios we mimic the effects of $\alpha$-enhancement by setting the stellar metallicity to Fe/H rather than the total metal content by mass, which largely traces the enrichment in oxygen.

\subsubsection{Stellar Rotation (MIST Isochrones)}\label{sec:mist}
Stellar rotation can have a profound impact on the properties of massive stars. Rotation-induced mixing impacts stellar lifetimes leads to greater mass loss and higher effective temperature \citep{maeder2000, Brott2011, Ekstrom2012, choi2016}. Rotating stars also produce a harder radiation field with more ionizing photons, which they can sustain for a longer duration due to the increased main-sequence lifespan \citep{byler2017}. This can be seen in Figure \ref{fig:Qh} where we consider an SSP of 1 $M_{\odot}$ and plot the time evolution of the number of ionizing photons emitted by the source (left panel) and the hardness of the radiation field (right panel) for different models. We use the ratio of the number of photons that can doubly ionize oxygen ($h\nu > 35$ eV) to the number of H-ionizing photons ($h\nu > 13.6$ eV) is as a proxy for radiation field hardness. To test the effects of stellar rotation we make use of \textsc{mist} stellar evolutionary tracks that come prepackaged with  \textsc{fsps}. These tracks were computed using the Modules for Experiments in Stellar Astrophysics (\textsc{MESA}) code \footnote{\url{https://waps.cfa.harvard.edu/MIST/}} \citep{dotter2016, choi2016, Paxton_2011, Paxton_2013, Paxton_2015}) and they include the effects of stellar rotation. 

An important point to note is that including stellar rotation-induced mixing can dredge up material from the stellar core up to the surface changing elemental production and mass-loss rates \citep{maeder2000}. \citet{roy2021} showed that stellar rotation can lead to higher rotational velocity leads to a decrease in [N/O] and an increase in [C/O] ratio. This can modify the metal abundances to have non-solar ratios, which can have a substantial impact on the emission-line ratios as discussed in \secref{sec:no_ratio}. This is, however, not included in our models as our photoionization calculations are conducted in post-processing.  This represents a slight inconsistency in our modeling. 

\subsubsection{Binary stars (\textsc{bpass})}\label{sec:bpass}
Binary stellar evolution is relatively common for massive stars. Recent estimates have shown that the binary fraction is around 70 - 90 \% for O and early-B type \citep[and references therein]{Mason2009, Chini2012, Sana2013, Sana2014, Gaspard2013} and around 20 - 40 \% for F and G type stars \citep{Raghavan2010, Tokovinin2014, Gao2014}. Having a stellar companion can substantially alter a star's evolution and drastically change the ionizing photon production \citep{wilkins2016}. This happens mainly through mass transfer and tides that open up evolutionary pathways that would otherwise be inaccessible. To model binary interactions we make use of v2.2 of Binary Population and Spectral Synthesis (\textsc{bpass}) model grids as described in \citet{eldridge2017} and \cite{stanway2018} with a Chabrier IMF and a cutoff of 100 $M_\odot$. We return to Figure~\ref{fig:Qh}, where we now highlight the hardness of the radiation field for \textsc{bpass} models as compared to our fiducial stellar models. The \textsc{bpass} model (shown in green) produces a substantially harder radiation field at ages above 5 Myr. 

\begin{figure*}
	\includegraphics[width=\textwidth]{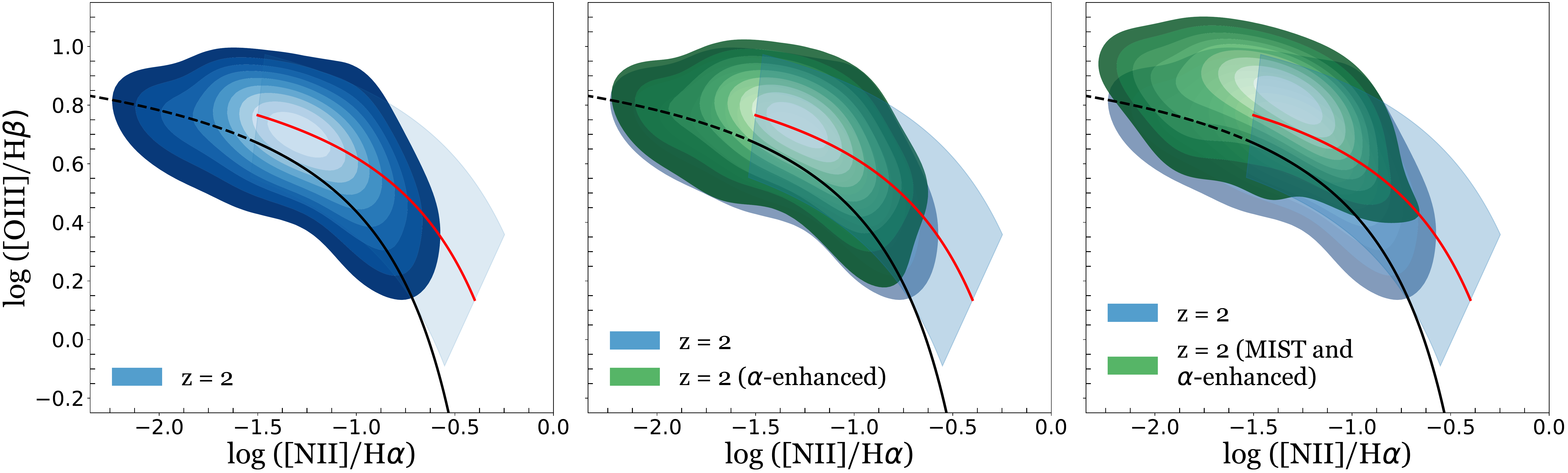}
    \caption {Effects of hardening the radiation field on the N2-BPT diagram.  Hardening the radiation field moves the curve towards higher O3 and lowers N2 thus, moving the distribution away from the observed sample. \textbf{Left panel:} This panel shows our base model at $z = 2$. \textbf{Middle Panel:} In this panel we overplot the galaxy distribution with stellar spectra hardening due to $\alpha$-enhancement taken into account (shown in green). As expected the galaxy distribution moves towards higher O3 although the increase just due to $\alpha$-enhancement is not substantial enough. \textbf{Right Panel:} Here, we show the effects of hardening the radiation field beyond that of just $\alpha$-enhancement by including both stellar rotation (MIST isochrones) and $\alpha$-enhancement (shown in green). As can be seen, this further hardening of the input spectrum leads to a substantial increase in O3 and a decrease in N2. Note: We see a similar increase in O3 when the radiation field is hardened by including the effect of binary stars through BPASS model grids, though do not show this for the sake of clarity.  In all the figures, SDSS and KBSS z $\sim$ 2.3 data points are approximated by polynomial fits adopted from \citet{kewley2013a} and \citet{strom2017}, defined by equation \ref{eq:locus_eqn1} (black line) and \ref{eq:locus_eqn2} (red line) respectively. The dashed black line shows the extrapolation of the SDSS locus beyond the SDSS-DR8 data points. The blue shaded region represents the intrinsic dispersion of $\pm$0.18 dex relative to the best fit curve (red line).}
    \label{fig:z2_ae}
\end{figure*}

\subsubsection{The Effects of Harder Radiation Fields Taken Together}
To better understand how the line ratios vary when the stellar spectrum is hardened, we employ simplified single-zone \textsc{cloudy} models. We run the \textsc{cloudy} models over a grid in age (2 -- 8 Myr) , metallicity (log (Z/Z$_\odot$): -0.6 -- 0.2), and logU (-4.0 -- -1.0). Throughout all the runs the cloud properties like abundance, density, $R_{\rm inner}$ are kept constant and the only thing that varied is the input stellar spectrum based on the corresponding grid value. The two main properties that we are interested in looking at are the hardness of the radiation field and the number of ionizing photons (log$Q$). As before, the ratio of the number of photons that can doubly ionize oxygen ($h\nu > 35$ eV) to the number of H-ionizing photons ($h\nu > 13.6$ eV) is used as a proxy for hardness.

In Figure \ref{fig:ae_pcolor} we show how the O3 and N2 line ratios change with hardness and number of ionizing photons, respectively. We find that with increasing hardness of the stellar spectrum, O3 always increases irrespective of $Q$. N2, in contrast, decreases with hardness for high $Q$ but increases with hardness for a low $Q$. This bimodality happens because nitrogen is in a singly ionized state in N2. Thus, when the hardness is increased and there are a sufficient number of ionizing photons then most of the nitrogen goes into higher ionization states, which leads to a decrease in N2. That said, when the count of ionizing photons is low, increasing the hardness leads to more nitrogen being in the singly ionized state, increasing N2. 

In Figure~\ref{fig:z2_ae}, we run a series of numerical experiments in which we show this in action. Specifically, we first include the effects of including $\alpha$-enhancement (top-middle panel) on the base $z = 0$ model. We find that $\alpha$-enhancement leads to a slight increase in O3. That said, the effect is relatively small. To further increase the impact of hardening the radiation field, in the top-right panel of Figure~\ref{fig:z2_ae}, we additionally include the MIST isochrones (that include the effect of rotating stars) with $\alpha$-enhancement. This increased hardening of the radiation field leads to a substantial increase in O3 and a decrease in N2. We see a similar increase in O3 when the radiation field is hardened by including the effect of binary stars through BPASS model grids (though for the sake of clarity, we omit this from Figure~\ref{fig:z2_ae}). Overall, we find that hardening the radiation field at fixed O/H in our base $z = 2$ sample does not move the peak of the \textsc{simba} galaxy distribution toward the observed peak in the KBSS sample. We, therefore, conclude that a harder radiation field at fixed metallicity by itself cannot be a potential driver of the mismatch between the simulations and observations at high redshifts.

\section{Discussion}\label{sec:discussion}
\subsection{Our results in the context of other studies}
The high-redshift N2-BPT offset (i.e., the observation that $z=2$ galaxies lie in a space of higher O3 and N2 than $z=0$ galaxies) has been a topic of heavy debate over the past decade. In the literature the explanations for the offset broadly fall into three categories: (1) higher ionization parameter, (2) harder radiation fields and (3) higher N/O ratio. In what follows, we examine our results in the context of these literature explanations for the high-$z$ offset.  

Our principal finding is that the offset toward larger O3 and larger N2 can be explained by sample selection alone, and that deeper observations that probe lower-mass galaxies will result in an average BPT diagram similar to the $z\sim 0$ locus. That said, this is in tension with the results inferred from observations. For example, \citet{shapley2015}, \citet{strom2017} and \citet{Runco2021} have all inferred that lower-mass galaxies are increasingly offset from the $z \sim 0$ sequence as compared to higher mass galaxies. 

A common theme that will emerge in this discussion is that the most of the methods employed to study the offset in the literature make use of lookup tables that are interpolated over a grid of physical parameters like metallicity, ionization parameter etc \citep[e.g.][]{gutkin, steidel2016, hirschmann2017, strom2017, Sanders_direct2020, ceverino_2021}. Some recent studies like \citet{katz_2019} have tried to overcome the rigidity of using lookup tables by employing an emulator-like approach. In specific, \citet{katz_2019} used a random forest machine-learning algorithm to train their nebular emission model over a grid of physical properties. We, on the other hand, have adopted a completely different approach for calculating the nebular emission-line spectrum in for cosmological galaxy formation simulations by computing the nebular line emission directly on a particle-by-particle basis. In this section, we go into detail about the modeling method used in some of the studies and place our results in the context of the broader literature.

\paragraph{Varying ionization parameter:} Several studies \citep[e.g.][]{brinchmann2008, kewley2013a, Kashino_2017, hirschmann2017, bian2020} have concluded that one of the main drivers of the high-redshift N2-BPT offset is the increase in ionization parameter. \citet{brinchmann2008} and \citet{hirschmann2017} attributed the increase in ionization parameter to higher specific SFR, whereas \citet{Kashino_2017} attributed it to a potential higher star formation efficiency (SFE) in high-redshift galaxies. Our findings are in tension with these results.  As discussed in section \ref{sec:ip_dec}, we find on the contrary that decreasing the ionization parameter can instead move galaxies {\it along} the BPT arc, and not toward higher O3 and N2 simultaneously.

Understanding the origin of the differing conclusions is difficult, though it is important to note that the differences in modeling techniques are significant. For example, theoretical studies like \cite{hirschmann2017} used galaxies from cosmological zoom simulations and included nebular emission in post-processing using \textsc{cloudy} model grids. They consider each galaxy to be a collection of H\scalebox{0.9}{ II} regions following the prescription of \citet{charlot2001}. They also account for metal depletion on dust grains and radiation pressure on dust in their cloudy models. This is in contrast to our model, which models H\scalebox{0.9}{ II} regions on a particle-by-particle basis and considers them to be dust-free. Some of the aforementioned observational studies employ single H\scalebox{0.9}{ II} region models when fitting their observed emission-line ratios, whereas our model treats galaxies as ensembles of H\scalebox{0.9}{ II} regions. We defer a full study comparing derived physical properties from individual H\scalebox{0.9}{ II} regions versus models that consider an ensemble to future work.

 How plausible is it to have a decreased ionization parameter in high-redshift galaxies? Several studies have suggested the opposite: that high-redshift galaxies might have a {\it higher} ionization parameter than their local counterparts \citep[e.g.][]{Brinchmann2004, liu2008, hainline2009, erb2010, wuyts2012, nakajima2013, shirazi2014, hayashi2015, onodera2016, Kashino_2017}. From Equation \ref{eq:U_eqn2} we can see that the ionization parameter is directly proportional to the hydrogen density and number of ionizing photons. We have already shown that decreasing hydrogen density can move the galaxy distribution toward lower O3 and higher N2. That said, it is unlikely that high-redshift galaxies have a lower hydrogen density. The other option is varying the number of ionizing photons, which is naturally related to the total cluster mass as well as the fractional contribution of massive stars to a stellar cluster's mass. A bottom-heavy IMF can also lead to lower production of ionizing photons \citep{van2008, cappellari2012, Conroy2012}.

\paragraph{Higher N/O ratio:} Studies like those of \citet{steidel2014}, \citet{masters2014}, \citet{Jones2015}, \citet{shapley2015}, \citet{sanders2016}, \citet{cowie2016}, and \citet{bian2020} have suggested elevated N/O ratio at high redshifts can be the reason for the high-redshift offset. In contrast, some theoretical studies \citep[e.g.][]{hirschmann2017} have found that increasing N/O can lead to an elevated N2 at fixed O3. We show (\S~\ref{sec:no_ratio}) that an increase in N/O abundance ratio in high-$z$ galaxies can move the peak of the \textsc{simba} galaxy distribution closer to the observed KBSS sample. 

Do galaxies at high$-z$ have a larger N/O ratio than comparable mass galaxies at $z=0$?  The fundamental challenge is that it is hard to measure N/O and metallicity (O/H) for the same galaxy in such a way that they do not implicitly depend on each other. Some studies, such as \citet{steidel2016}, have found no variation in N/O for high-redshift galaxies.  At the same time, other studies have suggested that some high-$z$ galaxies may show signs of N/O enhancement.  For example, \cite{masters2014}, using a composite spectrum of 26 galaxies at z $\sim$ 1.5 and z $\sim$ 2.2, showed that they have a log(N/O) higher by approximately 0.4 dex compared to local galaxies. A similar result was reached by \citet{strom2018}, where they found an increase in log(N/O) in their sample of about 0.34 dex. On the other hand, studies like \citet{bian2020} and \cite{strom2017} have found the log(N/O) in high-redshift galaxies to be higher by only about 0.1 dex.

\paragraph{Harder radiation field:} Recent studies by \citet{strom2018}, \citet{Sanders_direct2020}, \citet{Topping2020a}, \citet{Topping2020b}, and \citet{Runco2021} have argued that a harder radiation field primarily due to $\alpha$-enhancement is the main driver of the observed N2-BPT offset at high redshifts. Our results are in tension with these studies: we find that the BPT offset can be explained just by sample selection effects. In section \ref{sec:harder_field}, we looked at the effect of having a harder radiation field as a potential explanation for the subtle mismatch in the peak of the galaxy distribution between simulations and observations. Similar to the observational studies, using \textsc{cloudy} models on an SSP we found that having a harder radiation field leads to an increase in O3 and also N2 if the flux of ionizing photons are sufficiently high (see Figure \ref{fig:ae_pcolor}). That said, when implementing this directly into our simulations, the ionizing flux was insufficient to drive significant increases in O3 and N2 in our model galaxies. If a harder radiation field is indeed the driver of offsets at high-$z$, then this may represent a shortcoming in our simulations.

Trying to understand why we reach a different conclusion from these studies is nontrivial due to the drastically different methodologies involved. The observational studies in essence all start from observations and then use \textsc{cloudy} lookup tables to find the best-fit parameters. Our models, in contrast, approach the problem from the other direction: we employ theoretical models on simulated galaxies to match the observed line ratios. Relating the two directly is nontrivial. That said, we note that one potentially important difference is that in our models, the gas-phase nebular abundances are coupled to the stellar abundances, which may impact our results \citep{steidel2016}. We defer the exploration of this effect in detail to a future study. 

\subsection{Caveats}\label{sec:caveats}
\begin{figure*}
	\includegraphics[width=\textwidth]{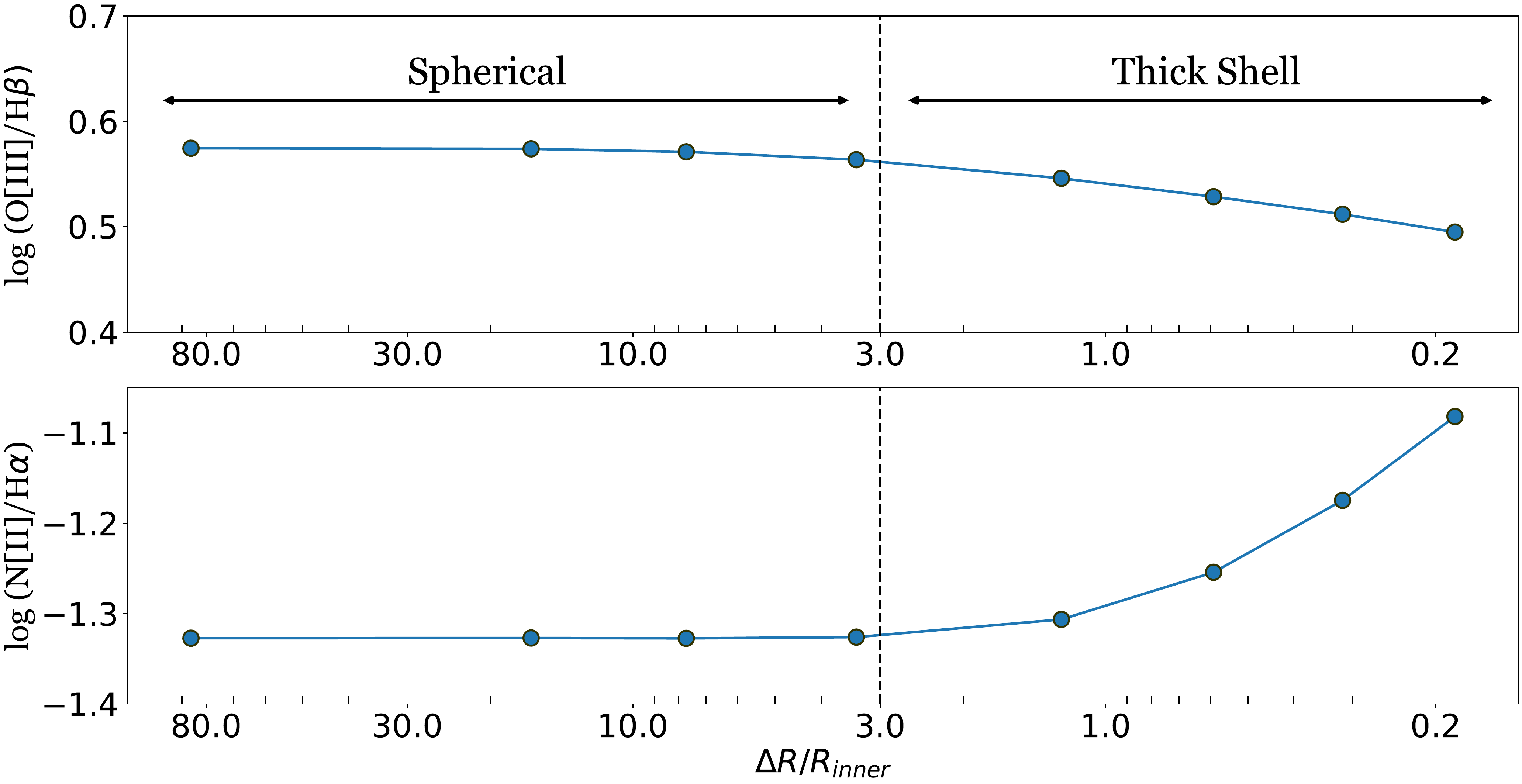}
    \caption{Variations in O3 and N2 line ratios as the H\scalebox{0.9}{ II} region geometry is modified. We find that uncertainty of about 0.2 dex and 0.1 dex in O3 and N2, respectively exists due to differences in H\scalebox{0.9}{ II} region geometry. The ratio of the thickness of the H\scalebox{0.9}{ II} region to the distance to the inner cloud boundary from the stellar source ($\Delta R$/$R_{\rm inner}$ ) is plotted on the x-axis. This ratio determines the geometry with the geometry being spherical if $\Delta R$/$R_{\rm inner} > $ 3 and $ 0.1 < \Delta R$/$R_{\rm inner} < 3 $. The point of transition ($\Delta R$/$R_{\rm inner}$ = 3) between the two geometries is shown by the dashed line. The top panel shows how the O3 line ratio varies as the H\scalebox{0.9}{ II} region gets thinner or as $\Delta R$/$R_{\rm inner}$ decreases. The bottom panel shows the same for the N2 line ratio.}
    \label{fig:geo}
\end{figure*}
In this section, we enumerate some caveats associated with our presented model. 

\subsubsection{H II region geometry}\label{sec:rinner}
In \textsc{cloudy} the H\scalebox{0.9}{ II} region geometry is determined by the ratio of the thickness of the H\scalebox{0.9}{ II} region ($\Delta R$) and the distance to the illuminated face of the cloud ($R_{\rm inner}$). It considered as spherical if $\Delta R$/$R_{\rm inner} >$ 3, a thick shell if  $\Delta R$/$R_{\rm inner} \leq $ 3 and plane parallel if $\Delta R$/$R_{\rm inner} < $ 0.1. As a reminder, in our fiducial analysis, we assume a spherical geometry. In this section, we test the effects of adopting different geometries on the nebular line ratios. To do this, we select a random star particle (age: $\sim$ 2 Myr, log($Z/Z_\odot$) = -0.03) from a model galaxy at $z=0$, and set up a grid of \textsc{cloudy} models. To get different geometries, we vary the $R_{\rm inner}$ from being very close to the star (spherical geometry) to it being comparable to $R_S$ (thick shell). 

In Figure \ref{fig:geo} we show how the line ratios change with $\Delta R$/$R_{\rm inner}$. The dashed line marks the point where $\Delta R$/$R_{\rm inner}$ = 3 which is the demarcation point between spherical and thick shell geometries. The line ratios stay more or less constant for the spherical geometry but they start to vary once the model transitions into thick shell geometry. 

The O3 decreases, whereas the N2 increases as the $\Delta R/R_{\rm inner}$ gets lower or as the H\scalebox{0.9}{ II} region gets thinner. This is because the [N\scalebox{0.9}{ II}] flux is mainly concentrated toward the outer edges of the H\scalebox{0.9}{ II} region while H$\alpha$ is emitted throughout the ionized region. Therefore, as the H\scalebox{0.9}{ II} region gets thinner and thinner, the volume of ionized gas emitting [N\scalebox{0.9}{ II}] compared to H$\alpha$ increases. The situation is exactly the opposite for [O\scalebox{0.9}{ III}] which is emitted closer to the inner boundary of the ionized region. This means as the H\scalebox{0.9}{ II} region gets thinner, the volume of ionized gas emitting [O\scalebox{0.9}{ III}] relative to H$\beta$ decreases, thus leading to a decrease in O3. 

In short, we find that an uncertainty of about $0.2$ dex in log([N\scalebox{0.9}{ II}]/H$\alpha$) and $0.1$ dex in log([O\scalebox{0.9}{ III}]/H$\beta$) exists due to difference in H\scalebox{0.9}{ II} region geometry. Because the physics that determines H\scalebox{0.9}{ II} region geometries are unresolved in our simulations, this represents a fundamental uncertainty in our modeling. 

\subsubsection{Other sources of ionizing radiation}
\paragraph{Diffused ionized gas (DIG)}
We model the H\scalebox{0.9}{ II} region as ionization bounded assuming that all the ionizing radiation is absorbed within the H\scalebox{0.9}{ II} region but, this is not true in reality. Some of the ionizing radiation leaks out and ionizes the surrounding medium leading to what is known as DIG. H\scalebox{0.9}{ II} regions can also form at the boundary of a molecular cloud causing them to poke out as they evolve leading to the formation of blister H\scalebox{0.9}{ II} regions. Studies like those of \citet{vale2019} and \citet{sanders2017biases} have argued that DIG can account for up to 50\% of the total nebular emission in local galaxies, and \citet{zhang2017} found that this increase can impact both O3 and N2.  The contribution from DIG is expected to be much more important for local galaxies as compared to their high-redshift counterparts. Even so, DIG might play an important role in understanding the N2-BPT offset therefore ignoring its effects can bias our findings. 
    
\paragraph{AGNs and post-AGB stars} 
We do not include the effects of AGNs and post-AGB stars. There is ionized gas present within the narrow-line region (NLR) of an AGN and in the envelopes of post-AGB and these sources can be a substantial contributor to the total nebular emission budget. The net effect of including them would be a harder ionizing radiation field because of the availability of more high-energy photons. 
    
\subsubsection{Sub-resolution effects}
In \textsc{simba} (as in any other large modern cosmological simulation) we cannot achieve the resolution required to model individual H\scalebox{0.9}{ II} regions. We are unable to probe the density and temperature regime of an H\scalebox{0.9}{ II} region and the Str\"{o}mgren radii of H\scalebox{0.9}{ II} regions are unresolved. This makes it necessary to include sub-grid physics in our models, which leads to uncertainties. As an example, the stellar feedback recipe may affect the composition and metallicity of the \textit{warm} gas component, depending on assumptions of how energy is ejected into and distributed around neighboring gas particles. Though this might not be as relevant for \textsc{simba}, where the winds are decoupled and explicitly deposit no energy into the ISM on their way out. Apart from this, in \textsc{simba} the sampling of the stellar ages of young star particles is rather stochastic. Finally, we are also missing dust physics within H\scalebox{0.9}{ II} regions. Future models will explore the impact of these aforementioned physical processes on the modeled nebular line emission from galaxies.

\subsubsection{N/O ratio}
As discussed in section \secref{sec:model_param} (see Figure \ref{fig:NO_z0}), the  \textsc{simba} z = 0 galaxies cannot reproduce the observed log(N/O) vs log(O/H) from \citet{pilyugin2012} (Equation \ref{pilyugin_eqn}). Due to this, we manually set the nitrogen abundance in our model such that Equation \ref{pilyugin_eqn} holds true; this manual tuning introduces a fundamental uncertainty in our model. Furthermore, although our our results (see \secref{sec:no_ratio}) and studies like \citet{masters2014, strom2017, strom2018} and \citet{bian2020} have shown that the N/O ratio at fixed O/H may evolve with redshift, we apply the \citet{pilyugin2012} relation when modeling  nebular emission at $z=2$ in order to avoid implementing an artificial redshift-dependent abundance variation when studying BPT offsets at high-$z$.

\section{Summary}\label{sec:summary}
We have developed a novel model for calculating the nebular emission-line spectrum on a particle-by-particle basis for cosmological galaxy formation simulations to investigate the observed offsets in N2-BPT space in high-redshift galaxies. The main results of our analysis are as follows.
\begin{itemize}
    \item We show that our model can successfully reproduce the observed SDSS-DR8 N2-BPT curve (\secref{sec:bpt_results}, Figure \ref{fig:z0}).  We also test our model against different mass bins, aperture sizes, and demonstrate that this correspondence is fairly robust (see Appendices \ref{sec:massbin} and \ref{sec:aperture_size}, Figure \ref{fig:bins} and \ref{fig:aperture}).
    
    \item We find that the N2-BPT offset as observed at high-$z$ arises primarily due to sample selection effects. The offset naturally appears in our simulation when we only consider the most massive galaxies. We predict that deeper observations of low mass galaxies at high redshifts with upcoming facilities like JWST will reveal that galaxies at high redshifts lie on the locus comparable to $z \sim 0$ observations  (\secref{sec:bpt_results}, Figure \ref{fig:z2}).  
    
    \item We find that even after considering sample selection effects there is still a subtle mismatch between simulations and observations: the peak of the galaxy distributions between simulations and observations do not agree. Via numerical experiments, we test a wide range of possible scenarios for driving the mismatch between simulations and observations, including varying ionization parameters (\secref{sec:ip_dec}, Figure \ref{fig:logU}), varying interstellar abundances in high-$z$ H\scalebox{0.9}{ II} regions (\secref{sec:no_ratio}, Figure \ref{fig:NO_ratio}) and,  harder radiation fields (\secref{sec:harder_field}, Figure \ref{fig:z2_ae}). We find that the most plausible explanation in our simulations is that either galaxies at $z \sim 2$ have increased N/O ratios or decreasing ionization parameter at fixed O/H as compared to the present epoch galaxies.
    
    \item In the future, we will explore the impact of additional physics, including nebular emission contribution from DIG, post-AGB stars, and AGNs. 

\end{itemize}

\section{Acknowledgements}
The authors would like to thank Chuck Steidel for providing KBSS data for our observational comparisons. \textsc{simba} was run using the DiRAC@Durham facility managed by the Institute for Computational Cosmology on behalf of the STFC DiRAC HPC Facility. The equipment was funded by BEIS capital funding via STFC capital grants ST/P002293/1, ST/R002371/1, and ST/S002502/1, Durham University, and STFC operations grant ST/R000832/1. R.D. acknowledges support from the Wolfson Research Merit Award program of the U.K. Royal Society.  P.G. and D.N. were funded by NSF AST-1909153.

\software{caesar \citep{Thompson2014}, cloudy \citep[v17.00][]{ferland2017}, cloudyfsps \citep{byler2017}, fsps \citep{conroy2009,conroy2010}, MESA \citep{dotter2016, choi2016, Paxton_2011, Paxton_2013, Paxton_2015}, powderday \citep{powderday}, python-fsps \citep{python_fsps}, simba \citep{simba}}

\bibliographystyle{aasjournal}
\bibliography{references}
\appendix
\section{Mass Bin Size}\label{sec:massbin}
\begin{figure}[htp]
	\includegraphics[width=\columnwidth]{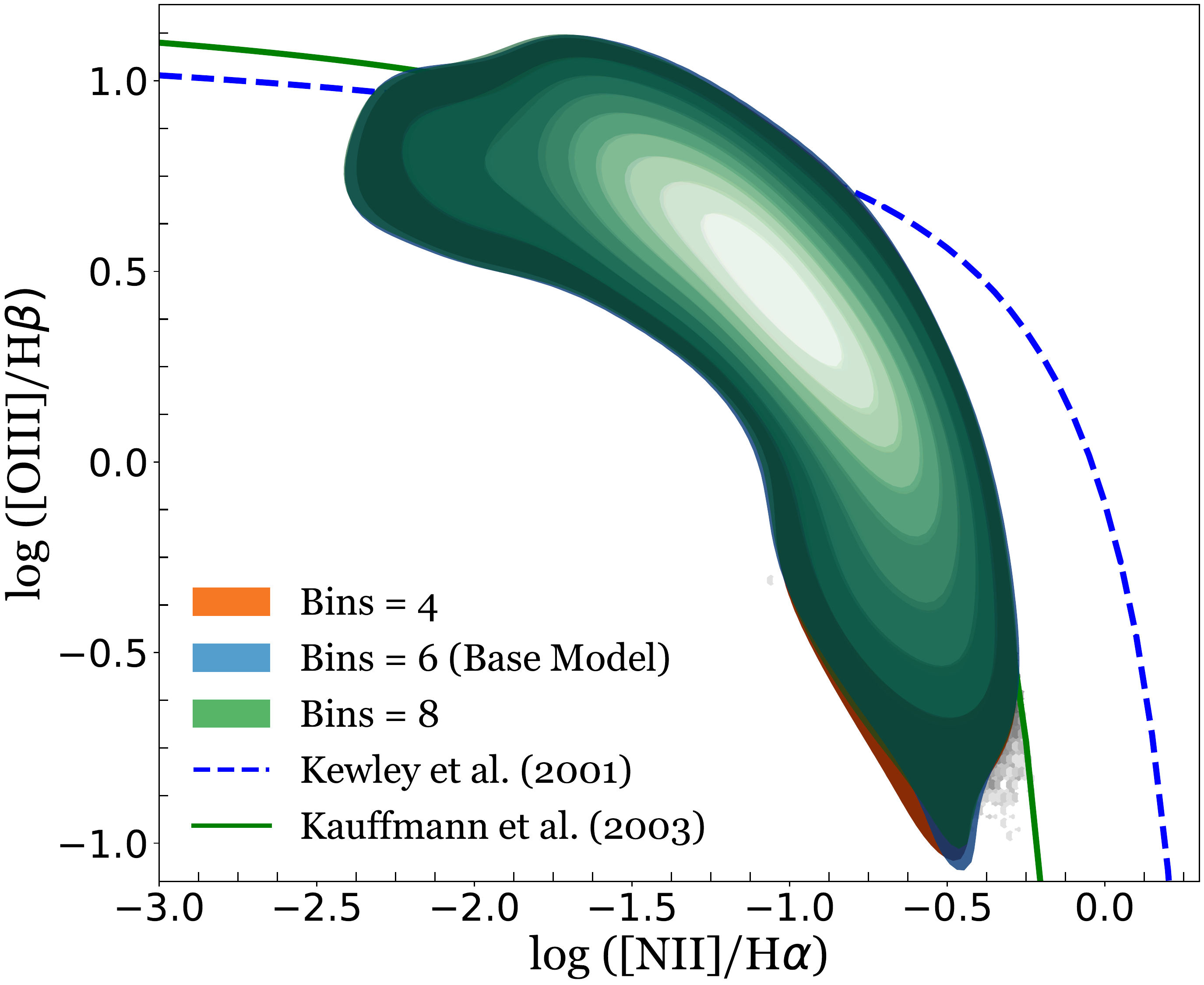}
    \caption{Effects of choosing different bin sizes for the cluster mass distribution. Results for bin sizes 4, 6, and 8 are shown in orange, blue, and green, respectively.}
    \label{fig:bins}
\end{figure}

As detailed in \secref{sec:model_param} we arbitrarily divide the cluster mass into 6 bins. To make sure our final result was not dependent on the bin size selected we ran the model again with a bin sizes of 4 and 8. Since we are using $\beta$ = -2.0 for the cluster mass distribution (see \secref{sec:method} Equation \ref{eq:cmdf_eqn}) it has the unique property that the total mass in each bin (Number of particles multiplied by the mass of single particle in that bin) is equal. This means that for a bin size of 4, 6, and 8 we have 1/4, 1/6, and 1/8 of the total mass in each bin respectively. Results are shown in Figure \ref{fig:bins} and as can be seen, our results are converged and do not depend on the bin size.

\section{Aperture Size}\label{sec:aperture_size}
\begin{figure}[htp]
	\includegraphics[width=\columnwidth]{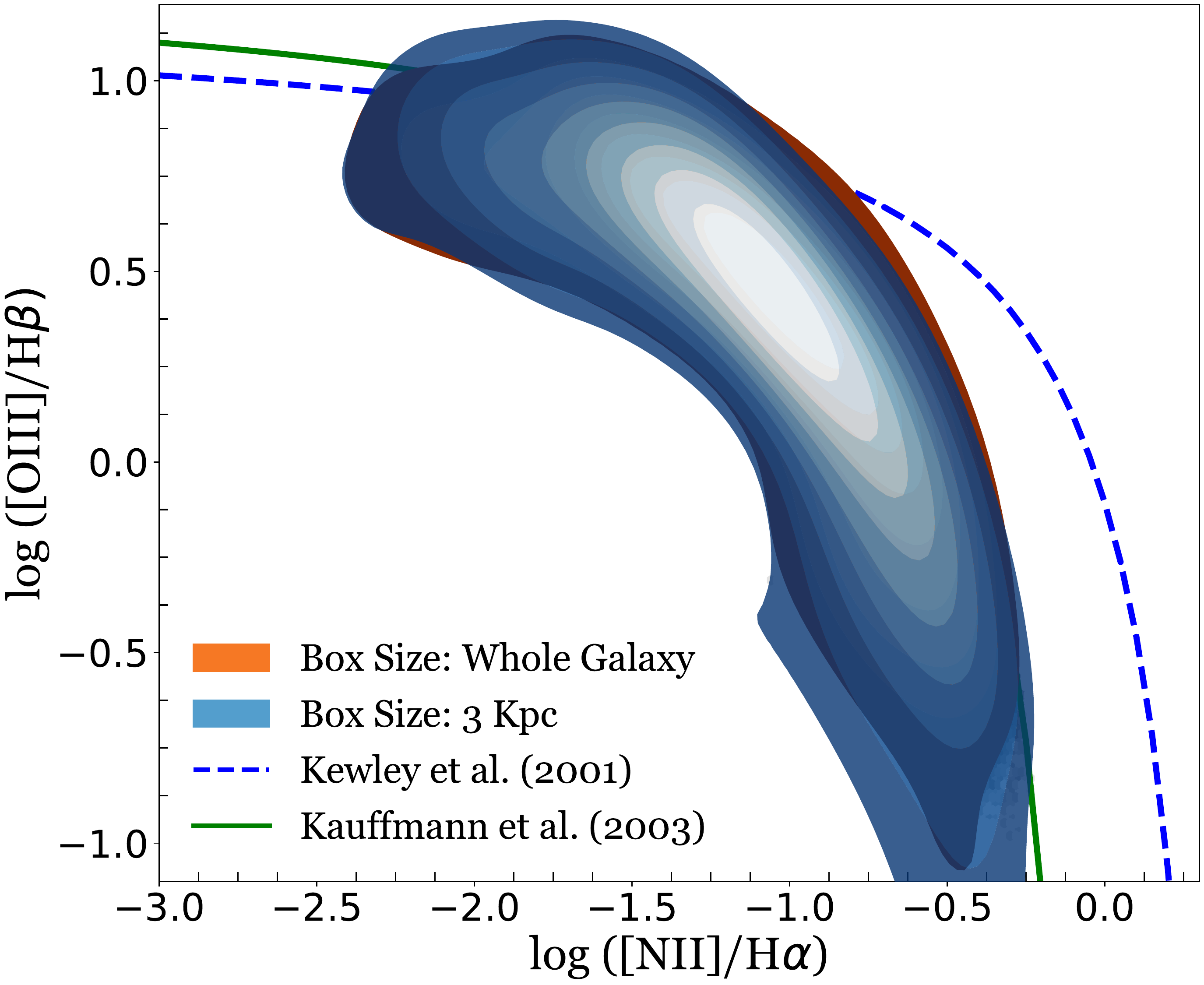}
    \caption{Effects of choosing different aperture box sizes for our model. The case when only particles within a central box of 3 kpc are taken into consideration is shown in blue and when all the particles in the galaxies are considered is shown in orange.}
    \label{fig:aperture}
\end{figure}

In the base model, we included nebular emission from all the young star particles (age $<$ 10 Myr) in the galaxy. On the other hand, owing to the small aperture size, SDSS effectively observes only the central few kiloparsecs of the galaxy. So to make sure we are doing an accurate comparison between our model galaxies and SDSS, we reran our models with this time only including the particles that lie within a box of 3 kpc around the center of the galaxy. As can be seen from Figure \ref{fig:aperture} the final result is pretty insensitive to what box size is being used. 

\end{document}